\documentclass[12pt]{article}
\usepackage[dvips]{graphicx}

\newcommand{\BH}	{{\rm BH}}
\newcommand{\weak}	{{\rm W}}
\newcommand{\planck}	{{\rm pl}}
\newcommand{\CP}	{{\rm CP}}
\newcommand{\univ}	{{\rm univ}}
\newcommand{\SM}	{{\rm SM}}
\newcommand{\bg}	{{\rm bg}}
\newcommand{\DW}	{{\rm DW}}
\newcommand{\sph}	{{\rm sph}}
\newcommand{\ReHeat}	{{\rm reh}}
\newcommand{\BBN}	{{\rm BBN}}
\newcommand{\DM}	{{\rm DM}}
\newcommand{\nucl}	{{\rm nucl}}
\newcommand{\crit}	{{\rm crit}}
\newcommand{\grav}	{{\rm grav}}
\newcommand{\peak}	{{\rm peak}}
\newcommand{\evap}	{{\rm evap}}

\newcommand{\formation}	{{\rm form}}
\newcommand{\reenter}	{{\rm ren}}
\newcommand{\lnear}{{\begin{array}{c} < \\[-0.8em] \sim \end{array}}}
\newcommand{\gnear}{{\begin{array}{c} > \\[-0.8em] \sim \end{array}}}

\newcommand{\TeV}	{{\rm \;TeV}}
\newcommand{\GeV}	{{\rm \;GeV}}
\newcommand{\MeV}	{{\rm \;MeV}}

\newcommand{\eV}	{{\rm \;eV}}
\newcommand{\kg}	{{\rm \;kg}}
\newcommand{\microg}	{{\rm \;\mu g}}
\newcommand{\tons}	{{\rm \;tons}}
\newcommand{\km}	{{\rm \;km}}

\newcommand{\K}		{{\rm \;K}}

\newcommand{\fig}[1]	{Figure \ref{#1}}

\begin{document}

\begin{flushright}
 \begin{minipage}[b]{43mm}
  YITP-01-28\\
  hep-ph/0104160\\
  April 2001
 \end{minipage}
\end{flushright}

\renewcommand{\thefootnote}{\fnsymbol{footnote}}
\begin{center}
 {\Large\bf
  Electroweak Domain Wall by Hawking Radiation:\\[0.5em]
  Baryogenesis and Dark Matter\\[0.5em]
  from Several Hundred kg Black Holes\\}
 \vspace*{3em}
 {\large Yukinori Nagatani}\footnote
 {e-mail: nagatani@yukawa.kyoto-u.ac.jp}\\[1.5em]
 {\it Yukawa Institute for Theoretical Physics, Kyoto University,\\
      Sakyo-ku, Kyoto 606-8502, Japan}
\end{center}
\vspace*{1em}

\begin{abstract}
 We show that a spherical electroweak domain wall
 is formed around a small black hole
 and this is a general property of the Hawking radiation
 in the vacuum of the Standard Model.
 The wall appears
 not only for the first order phase transition in the electroweak theory
 but also for the second order one
 because the black hole heats up its neighborhood locally
 by the Hawking radiation in any case.
 We propose a model for
 unifying the origin of the baryon number and the cold dark matter
 in our universe
 by using properties of the primordial black hole
 with a mass of several hundred kilograms.
 The interaction between our wall
 and the Hawking-radiated-particles
 can create a baryon number
 which is proportional to the mass of the black hole
 as well as the CP broken phase in the extension of the Standard Model.
 Our model can explain both the baryon-entropy ratio $B/S \sim 10^{-10}$
 and the energy density of the dark matter,
 provided that the following three conditions are satisfied:
 (i) the primordial black holes dominate in the early universe,
 (ii) the CP broken phase in the wall is in the order of one and
 (iii) any black hole leaves a stable remnant with a Planck mass
 after its evaporation.
 Our model also predicts a cosmological graviton background
 with a peak-energy $120 \sim 280$ eV in the present universe.
\end{abstract}

\newpage
\section{INTRODUCTION}\label{intro.sec}

A particle radiation from a black hole as a quantum effect
was predicted by Hawking in 1975 \cite{Hawking}.
The radiation has a thermal spectrum and
its temperature is known as the Hawking temperature.
In the case of a Schwarzschild black hole with a mass $m_\BH$,
the Hawking temperature is proportional to $m_\BH^{-1}$
and the luminosity of the radiation is proportional to $m_\BH^{-2}$,
then
a very small black hole has very violent radiation
into the very small region.
For example,
the luminosity of a black hole with a mass of seven tons
is comparable to the solar luminosity $1.60\times10^{12}\GeV^2$,
however, the Schwarzschild radius
as a size of the shining region is only $10^{-21}$ cm
which is $10^{-7}$ of the proton's radius.
%
%
Therefore, a neighborhood of the small black hole
provides a very interesting laboratory
for the particle physics and the field theory.
From this point of view, 
Cline {\it et.al.} discussed 
a formation of a QCD fireball around the black hole \cite{Cline}.
Heckler discussed a formation of a photosphere near the black hole
by QED interaction \cite{Heckler}.

\begin{figure}[htbp]%
\begin{center}%
 \ \includegraphics{BHWall.eps}%
 \caption{
 Structure of
 the spherical electroweak domain wall around a black hole.
 A black hole with a mass $76\kg \sim 200 \tons$
 can heat up its neighborhood by the Hawking radiation
 and restores the electroweak symmetry in the neighborhood spherically.
 The absolute value of the Higgs VEV $|\left<\phi\right>|$
 is depending on the distance from the black hole $r$.
 The shaded (magenta) region, where the Higgs VEV is varying,
 is a spherical electroweak domain wall surrounding the black hole.
 In this figure, we have assumed
 that the electroweak phase transition is the second order. 
 The radius of wall $r_\DW$ and the width of wall $d_\DW$
 is almost same, however,
 the Schwarzschild radius $r_\BH$ is much smaller than those values.
 The concrete values of the parameters are shown in \fig{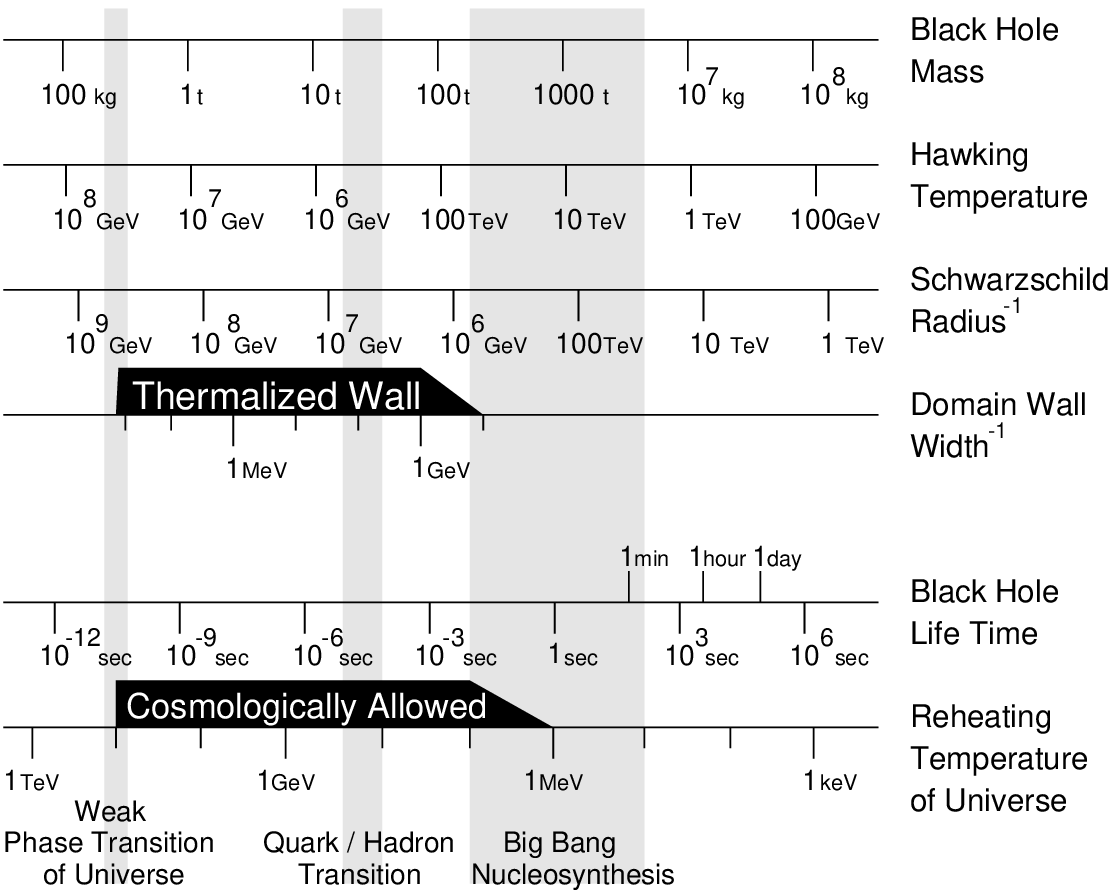}.
 }%
 \label{BHWall.eps}%
\end{center}%
\end{figure}

Then it is very interesting
to consider the Hawking radiation of the black hole
within the framework of the Standard Model (SM),
especially in the physics of the electroweak theory.
%
%
When we put a small black hole into the electroweak broken phase vacuum,
we can easily imagine that
the black hole heats up its neighborhood by the Hawking radiation 
and
it restores the electroweak symmetry
in the near region around the black hole.
Therefore we expect a formation of the spherical electroweak domain
wall surrounding the black hole,
which separates the broken phase region and the symmetric phase region
(see \fig{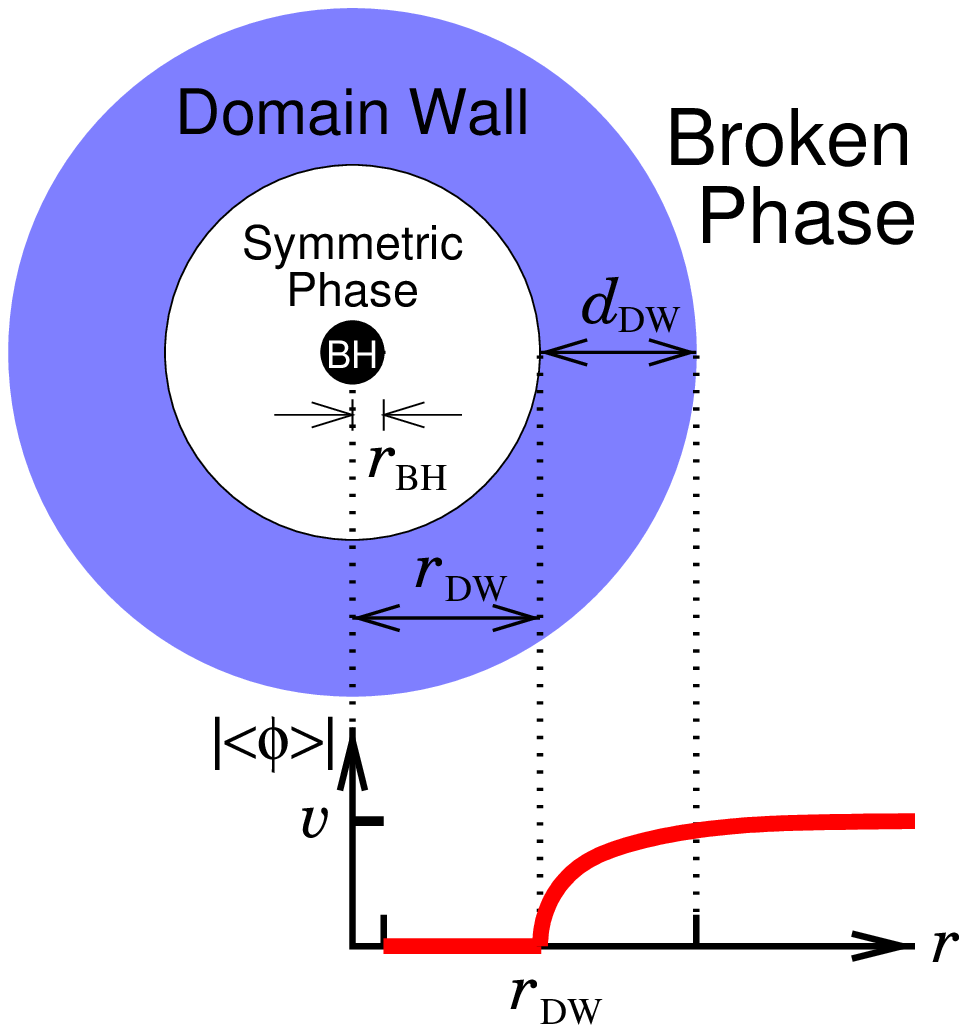}).
We have discussed such a subject
by considering 
a thermalization of the SM-particles radiated from the black hole
by the interactions of the SM \cite{Nagatani}.
Consequently
we have shown an appearance of the domain wall
as a general property of the black hole in the SM background
when the condition $76\kg \lnear m_\BH \lnear 200 \tons$
is satisfied.
It is an important point
that the formation of our domain wall does not require the first order phase
transition of the electroweak theory.
In the ordinary situation of a phase transition with second order,
any domain wall which separates the ordered and disordered phase regions
does not appear,
however, the Hawking radiation can heat up the neighborhood locally,
therefore we can find the wall even in the second order case.
In the \fig{BHWall.eps},
we have shown the resultant domain wall
in the case of the second order phase transition.

Formation mechanisms of the small black holes in the early universe
have been considered 
by the density fluctuations \cite{Formation,CarrForm},
fluctuations from the inflations
\cite{FormationByInflation, FormationAndDarkMatter, CarrLimit, Hsu},
and the first order phase transitions \cite{FormationPhaseTrans}.
These black holes created in early universe are called
``primordial black holes''.
As a cosmological consequence
the primordial black holes may contribute to
the nucleosynthesis \cite{BHBBN},
the dark matter \cite{FormationAndDarkMatter},
the baryogenesis \cite{Barrow} and so on.
Barrow {\it et al.} proposed a baryogenesis scenario
by the primordial black holes based on the grand unified theory (GUT)
\cite{Barrow}.
They consider that 
the primordial black holes with $T_\BH > 10^{14} \GeV$
radiate GUT-bosons and their decay creates a baryon number.
Their scenario is a kind of the GUT-baryogenesis scenario \cite{GUTBG},
then these theory cannot avoid
the washout problem by the sphaleron process \cite{Sph}.
To avoid this problem,
Majumdar {\it et al.} consider the GUT-baryogenesis by black holes
survived in the symmetric phase vacuum \cite{Majumdar}.
A natural solution for the problem
may be the ``electroweak baryogenesis''
proposed by Cohen, Kaplan and Nelson \cite{CKN,CKN2},
their scenario is frequently called CKN model.
They applied the sphaleron process
not to wash out bat to create the baryon number.
In the CKN model,
the first order phase transition of the electroweak theory
is assumed
and the formation and the running of the electroweak domain wall
in the universe at the critical temperature plays a crucial role.

Here, we can ask if the domain wall by the Hawking radiation
can create baryon number by using a mechanism in
the electroweak baryogenesis scenario.
We have discussed such a subject by a rough estimation \cite{NagataniOld}
and more precise analysis \cite{Nagatani}.
We have shown
that a system consisting of a black hole and our domain wall
satisfies the Sakharov's three criteria \cite{Sakharov}
and have concluded that the interaction between the wall and the Hawking
radiated particles create baryon number
when the wall have a CP broken phase.
The total created baryon number $B$
by a evaporation of a black hole with a mass $m_\BH$
is given by very simple relation:
\begin{eqnarray*}
 B &=& 10^{-9} \times \Delta\theta_\CP \; \frac{m_\BH}{T_\weak},
\end{eqnarray*}
where $\Delta\theta_\CP$ is the CP broken phase in the wall
and $T_\weak \simeq 100\GeV$ is the critical temperature of the
electroweak phase transition.
As well as the CKN model,
we should assume the CP broken phase in the wall
and this requires an extension of the SM,
because the CP broken phase in the yukawa sector
of the SM (the Kobayashi Maskawa phase) \cite{KM}
is too small and not suitable for our purpose.
Therefore we need an extension of the SM.
In the CKN model the SM with two Higgs doublet extension (2HSM) is assumed
as a simplest model for the additional CP broken phase.
We also assume the 2HSM as a back ground field theory of the black hole.

Now, we know that an evaporation of a small black hole
can create a baryon number,
then one would like to apply the phenomena to the cosmology, i.e.,
a electroweak baryogenesis scenario by the primordial black holes.
We find out that
the origin of the baryon number in the universe can be explained
by assuming (i) the very early universe
earlier than the age $t_\univ \sim 10^{-10} \ \sec$ is dominated by
primordial black holes with a mass of several hundred kilograms:
\begin{eqnarray*}
 m_\BH
  &\simeq&
  \frac{1}{20} \left(\frac{m_\planck}{T_\weak}\right)^{2/3} m_\planck
\end{eqnarray*}
and (ii) the CP broken phase in the wall $\Delta\theta_\CP$ is order one.
Namely, our scenario results the baryon-entropy ratio $B/S \sim 10^{-10}$
which is required from the observation
and prediction from the big-bang nucleosynthesis (BBN) theory
\cite{KolbTurner}.
The merit of the scenario is that
the scenario does not require
the assumption of the first order phase transition.
However, the scenario require
the assumption of the black hole dominated universe.
Our scenario is summarized in \fig{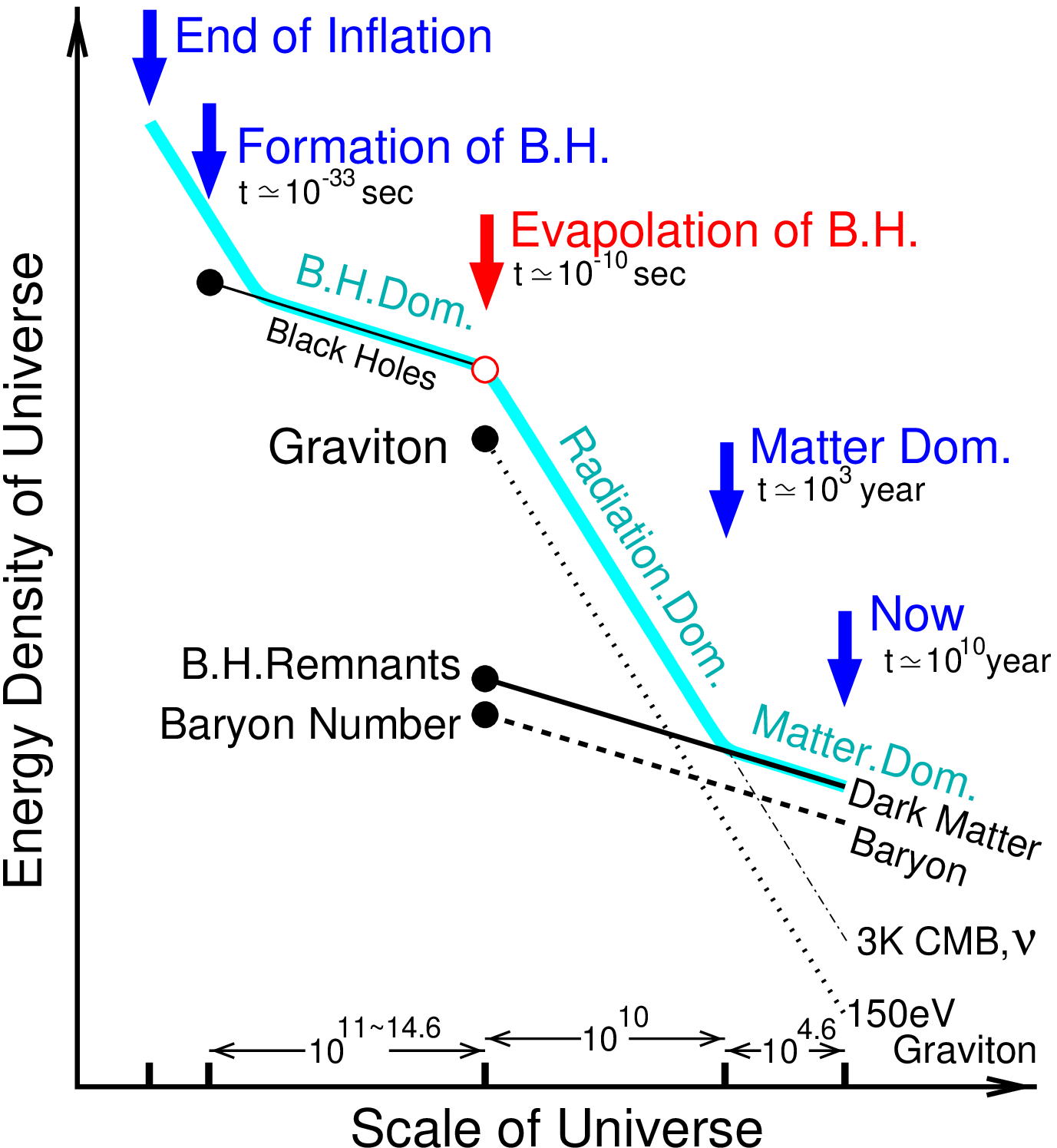}.

\begin{figure}[htbp]%
\begin{center}%
 \includegraphics{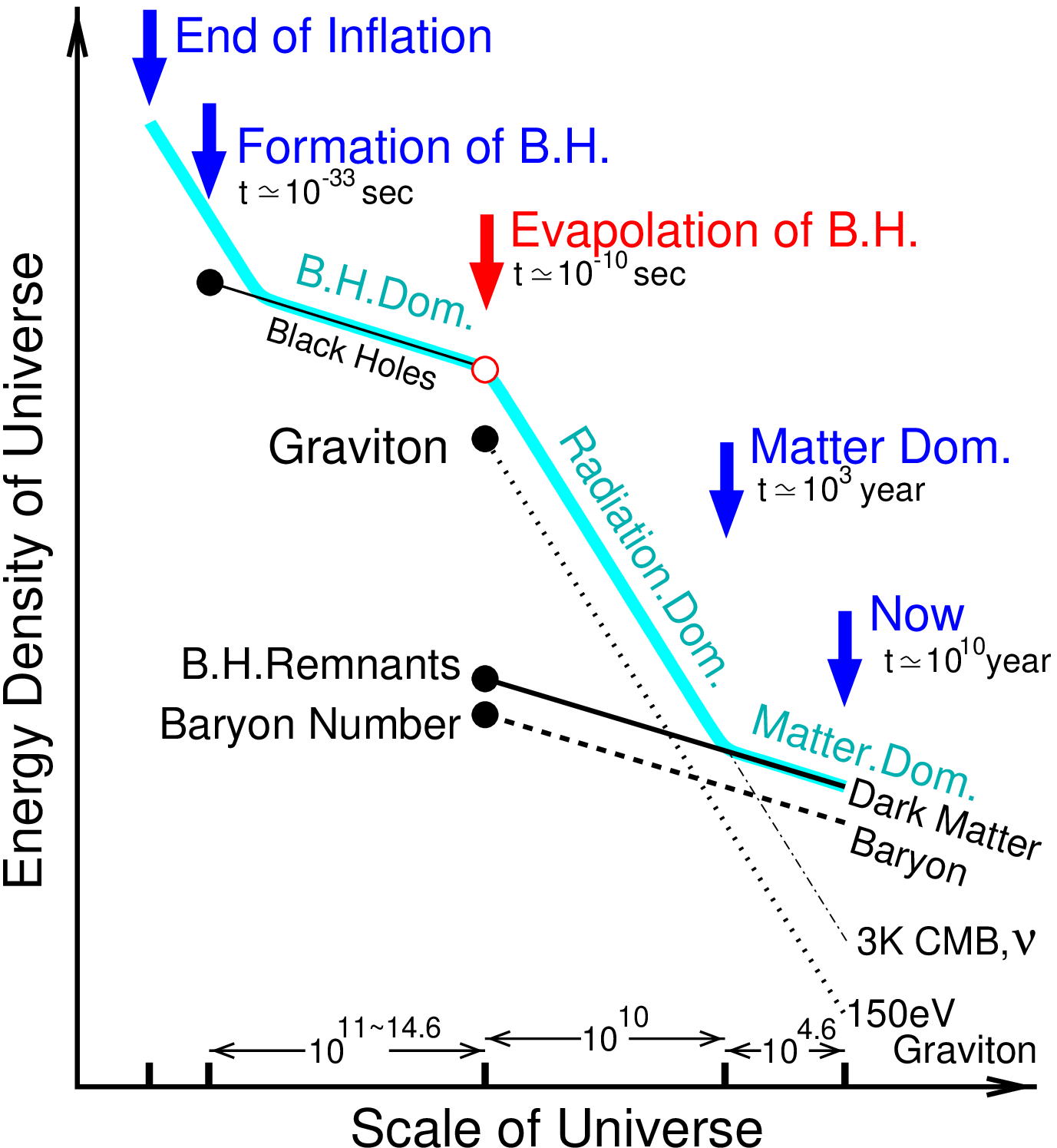}%
 \caption{
 History of the energy densities in the universe
 from the end of inflation to now in our model.
 Both axes are logarithmic.
 The thick (cyan) curve describes the energy density dominating
 the right hand side of the Einstein equation.
 The primordial black holes, the remnants of the black holes
 and the baryon numbers
 are dust like matters, and their densities obey $R^{-3}$
 to the scale of the universe $R$.
 Any radiations including the gravitons and the photons obey $R^{-4}$.
 The baryon number, the dark matter as the black hole remnants,
 the graviton background
 and ordinary radiation consisting of the SM particles
 are created by the evaporation of the black holes
 at $t \simeq 10^{-10}\sec$.
 In our model,
 the universe after the inflation develops four eras;
 the radiation dominant era reheated at the end of the inflation,
 the black hole dominant era, the (ordinary) radiation dominant era
 reheated by the black holes
 and the (ordinary) matter dominant era.
 The history after the evaporation of the black holes in our model
 is the same as the history of the standard cosmology.
 }%
 \label{DHistory.eps}%
\end{center}%
\end{figure}

Several authors considered 
what is left after the completion of a black hole's evaporation
\cite{Zeldovich, PBHFormation},
which is referred as the remnant or the relic of a black hole.
The simple assumption is that no remnant is left,
however, the issue should be considered
by the theory of quantum gravity which does not have been understood.
For example,
Zel'dovich discussed that
any black hole leaves a stable remnant with a mass of the Planck scale
$\simeq 20 \microg$ (Planck remnant)
because of the uncertainty principle for the mass of the black hole and
the Schwarzschild radius \cite{Zeldovich}.
If primordial black holes leave remnants,
they should constitute one of the cold dark matters
\cite{MacGibbon, RemnantCosmology}.
In our model,
we have assumed the domination of the black holes
with a mass of several hundred kilograms
to create baryon numbers,
therefore, the assumption for the Planck remnants implies
that they exist in the present universe and have a energy density.
We have estimated that the energy density of the remnants is
about ten times as large as the density of the baryonic matter,
namely,
the energy density of the remnants
is close to the critical density of the universe.
Therefore,
we can consider that the remnants of the black holes in our model
are main contents of the cold dark matter.

The Hawking radiation contains particles of all species in the SM
and also contains the gravitons
because the gravitational interaction is universal.
The assumed primordial black holes should radiate the gravitons
with the thermal spectrum,
however,
the radiated gravitons in the universe are not thermalized
because the interaction of the gravitons is especially small.
The Hawking temperature continues to rise until the evaporation ends,
therefore the energy spectrum of the graviton in the universe
should not be thermal.
We have evaluated the spectrum and the energy density.
In the present universe,
the spectrum of the gravitons has a peak at $120 \sim 280 \eV$,
and the energy density is about $1/82$ of
the cosmological microwave background (CMB) with $2.75\K$.
This gravitons should be called the cosmological graviton background (CGB).
The standard cosmology
with considering the graviton decoupling at the Planck scale
predicts the CGB with thermal spectrum of $0.91\K$,
whose energy density is about $1/82$ of the CMB
and whose spectrum has peak at $0.079\ \rm meV$ \cite{KolbTurner}.
The energy density of CGB in our scenario
equals that in the standard cosmology,
however,
the peak energy of the CGB spectrum in our scenario
is much different from that in the standard cosmology.

\begin{figure}[htbp]%
\begin{center}%
 \ \includegraphics[scale=1.35]{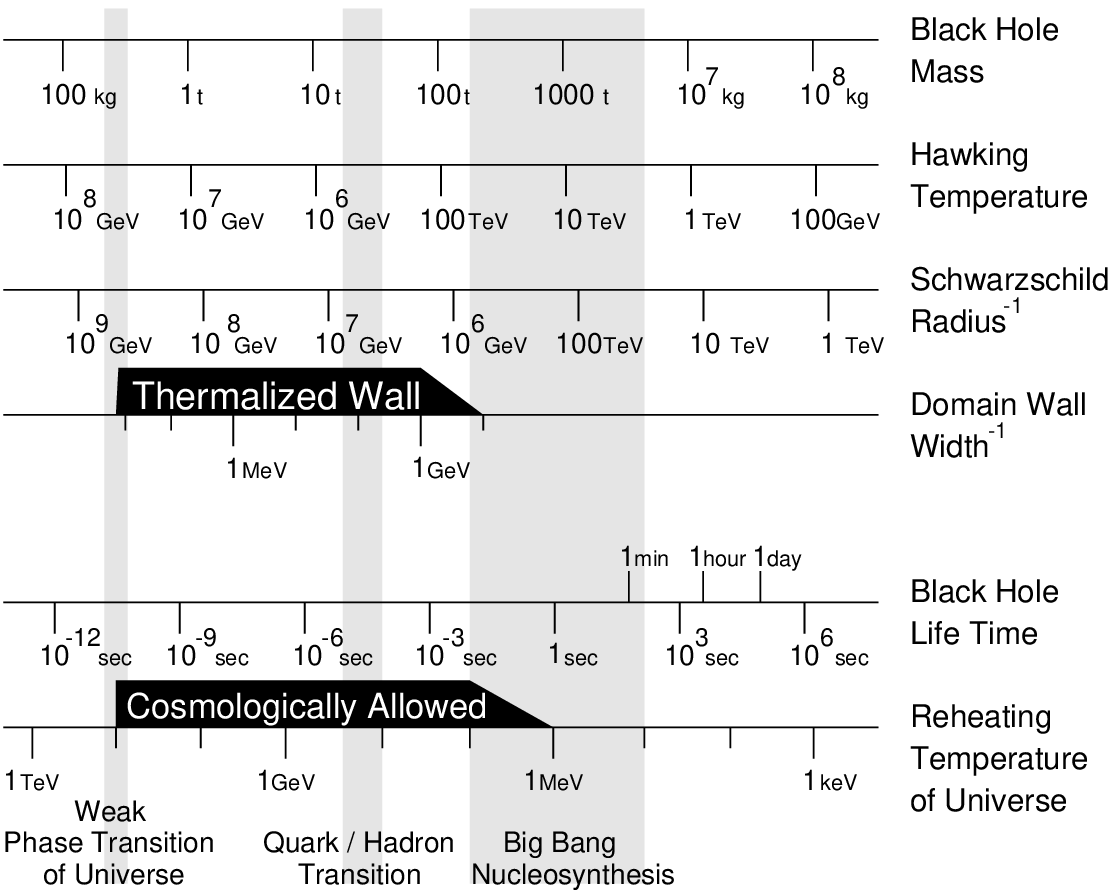}%
 \caption{%
 The relations among parameters for the Schwarzschild black hole,
 the formated electroweak domain wall and the universe in our model.
 In the parameter region marked with ``{\sf Thermalized Wall}'',
 the stationary thermalized wall appears
 and its width is given by the inverse of the number straight line
 ``$\sf Domain\ Wall\ Width {}^{-1}$''.
 The bottom number straight line describes the temperature of
 the universe reheated by the black holes just after the evaporation
 in our model.
 The parameter region marked with ``{\sf Cosmologically Allowed}''
 is allowed parameter region cosmologically, namely,
 the reheating temperature should be lower than the electroweak critical
 temperature to create a wall, and the Hawking radiation should not prevent
 the mechanism of the Big-Bang Nucleosynthesis.
 }%
          \label{BHTemp.eps}%
\end{center}%
\end{figure}

In this paper, we consider the Hawking radiation from
the Schwarzschild black hole only which is parameterized by a parameter.
We note relations among the mass $m_\BH$, the Hawking temperature $T_\BH$,
the Schwarzschild radius $r_\BH$ and the lifetime $\tau_\BH$:
\begin{eqnarray}
 T_\BH		&=&  \frac{1}{8 \pi} \frac{m_\planck^2}{m_\BH},\\
 r_\BH		&=& 2 \frac{m_\BH}{m_\planck^2}
 		\;=\; \frac{1}{4 \pi} \frac{1}{T_\BH},\\ 
 \tau_\BH	&\simeq&  \frac{10240}{g_*} \frac{m_\BH^3}{m_\planck^4}
		\;=\; \frac{20}{\pi^2 g_*} \frac{m_\planck^2}{T_\BH^3},
\end{eqnarray}
where $g_*$ is the freedom of the massless particles that
the black hole can decay into at its temperature.
%
We display these relations with our results in \fig{BHTemp.eps}.
We will parameterize black holes
by either the mass $m_\BH$ or the Hawking temperature $T_\BH$
which allows us more simple description.

The paper is organized as follows;
%
In Section \ref{DW.sec}
the thermalization of the SM particles radiated from a black hole
by the SM interactions is discussed
and the formation of a spherical domain wall is shown.
%
In section \ref{Baryon.sec}
we discuss the Sakharov's three criteria for the baryon-number-creation
in the domain-wall-system around the black hole
and we obtain a description for
how much baryon number is created by a black hole.
%
In section \ref{Cosmology.sec}
we provide a model for baryogenesis by the primordial black holes.
%
In section \ref{DM.sec}
we evaluate
the energy density of the remnants of the black holes
for our baryogenesis scenario
and 
we discuss that we can regard the remnants as a cold dark matter
in the present universe.
%
In section \ref{model.sec}
we calculate the energy spectrum of the graviton back ground (CGB)
in the present universe
which is producted by the evaporation of the black holes.
%
In section \ref{form.sec}
we discuss
the requirement to the inflation model
for formating the primordial black holes in our model.
%
In section \ref{summary.sec}
we provide a conclusion and discussions.

\section{Formation of Spherical Domain Wall}\label{DW.sec}

In general, 
a local thermalization is realized
when a local part of the system has interactions enough.
More precisely,
a local thermalization is realized
when a mean free path for particle $\lambda$ is smaller than
the length scale of the local system.
When we can make sure the local thermalization,
we can define local temperature $T(x)$ approximately
which depends on the position $x$.
This is a local thermal equilibrium (LTE) approximation.
To confirm a local thermalization
near a black hole by the Hawking radiation
and to obtain a local temperature distribution
by the matter transfer phenomenon
within the framework of the Standard Model (SM),
we should consider the mean free paths for the SM-particles
by the interactions of the SM.

A mean free path for a particle $f$ is given by
\begin{eqnarray}
 \lambda_f &=& \left[ \sum_F n_F \; \sigma_{F,f} \right]^{-1},
 \label{MFP}
\end{eqnarray}
where $f$ and $F$ denote a species of the particles
and $\sum_F$ is a summation all over the particle species
in the Standard Model.
We will consider a (local) thermal equilibrium
of the electroweak symmetric plasma with a temperature $T > T_\weak$.
A number-density of the particle $F$ is given
by the thermodynamical formula:
\begin{eqnarray}
 n_F &\equiv& \frac{\zeta(3)}{\pi^2} g_{*F} T^3,
\end{eqnarray}
where a constant $g_{*F}$ means a massless freedom for particle $F$.
A cross section between particle $f$ and $F$ is approximately given by
\begin{eqnarray}
 \sigma_{f,F} &\equiv& \frac{\alpha_{f,F}^2}{T^2}
\end{eqnarray}
where $\alpha_{f,F}$ means dimensionless coupling constants.
By combining these equations,
we obtain simple form for the MFPs:
\begin{eqnarray}
 \lambda_f(T) &=& \frac{\beta_f}{T}, \label{MFP.eq}
\end{eqnarray}
where $\beta_f$ is a constant depending on the particle species only.

The particles in the plasma with the symmetric phase
of the Standard Model are divided into three classes
by the value of $\beta_f$.
First, the quarks and the gluons have a strong interaction $\alpha_s$ and
they have the shortest MFP: $\beta_s \simeq 10$.
Second, the left handed leptons $l_L, \nu_L$
and weak bosons $W^\mu$ have a $SU(2)_W$-interaction $\alpha_W$
and do not have the strong interaction,
they have a middle MFP $\beta_W \simeq 100$.
Finally, the right handed leptons $l_R$ and $U(1)_Y$-gauge bosons
have the only interaction of the hypercharge $\alpha_Y$,
therefore they have the longest MFP: $\beta_Y \simeq 1000$

Now, we consider a black hole whose Hawking temperature $T_\BH$ is
much greater than the electroweak critical temperature $T_\weak$.
From the form of the MFP in the equation (\ref{MFP.eq}),
we can conclude that the closely neighborhood of the black hole
is not thermalized
because the Schwarzschild radius is much smaller
than the shortest MFP in the Hawking temperature $T_\BH$:
\begin{eqnarray}
 r_\BH =
  \frac{1}{4\pi} \frac{1}{T_\BH}
  &\ll&
  \lambda_\BH =
  \beta_f \frac{1}{T_\BH},
\end{eqnarray}
where the shortest MFP is given by $\beta_s \simeq 10$
for the quarks and the gluons.
The fact of the non-thermalization in the closely neighborhood
is important,
because the fact ensures that
the black hole can freely Hawking-radiate the particles.
If the closely neighborhood is thermalized by Hawking radiation,
many of the radiation particles return to the black hole,
namely,
flux of the Hawking radiation does not obey the Stefan-Boltzmann's law.

The non-thermalization in the closely neighborhood
does not mean that there are no thermalized regions.
When the local thermalization by the Hawking radiation is realized,
the local temperature distribution $T(r)$
should be a function of distance from the center of the black hole $r$
because of its spherical symmetry.
The Hawking radiation initially should thermalize
a sphere with the radius
\begin{eqnarray}
 r_s &=& \lambda_s(T(r_s))
\end{eqnarray}
to the temperature $T_s = T(r_s)$.
We call the radius $r_s$ a minimal thermalized radius
and we call the temperature $T_s$ a initial temperature.
Later we determine this radius $r_s$ and the temperature $T_s$.
For every sphere greater than the minimal thermalized radius,
we can make sure of the thermalization of such a sphere
by the temperature distribution $T(r)$.

To determine the temperature distribution $T(r)$,
we discuss the transfer equation of the energy.
The energy diffusion current for the particle species $f$
in the LTE approximation is
\begin{eqnarray}
 J_{f,i}(T(x)) &=& - D_f(T(x)) \; \partial_i \rho_f(T(x)),
\end{eqnarray}
where
\begin{eqnarray}
 D_f(T(x)) &=& \frac{1}{3} \lambda_f(T(x))
  \;=\;
  \frac{1}{3} \frac{\beta_f}{T(x)},
\end{eqnarray}
is diffusion parameter.
The energy density for particle species $f$ is 
\begin{eqnarray}
 \rho_f(T(x)) &=&
  \frac{\pi^2}{30} g_{*f} T^4(x)
\end{eqnarray}
The total energy density and total diffusion current are
given by sum of energy density
for all particle species in the Standard Model:
\begin{eqnarray}
 \rho &=& \sum_f \rho_f \;=\; \frac{\pi^2}{30} g_{*\SM} T^4(x),\\ 
 J_i &=& \sum_f J_{f,i}
  \;=\;
  - \frac{\pi^2}{90} g_{*\SM} \frac{\beta_\SM}{T(x)} \: \partial_i T^4(x),
\end{eqnarray}
where we have defined $g_{*\SM} \equiv \sum_f g_{*f} = 106.75$ as
a massless freedom for all particles in the Standard Model
and have defined
\begin{eqnarray}
 \beta_{\SM}
  &\equiv& \frac{1}{g_{*\SM}} \sum_f \beta_f g_{*f}
  \;\simeq\; 100
\end{eqnarray}
as a effective $\beta$-constant for the MFP.
The definition of the energy transfer equation is
only a equation for the energy preservation:
\begin{eqnarray}
 \frac{\partial}{\partial t} \rho &=& - \nabla_i J^i,
\end{eqnarray}
this treatment is referred to as the diffusion approximation of
photon transfer at the deep light-depth region \cite{Mihalas}.
We can easily obtain 
a stationary spherical-symmetric solution for the transfer equation:
\begin{eqnarray}
 T(r) &=& \left[ T_\bg^3 + (T_s^3 - T_\bg^3) \frac{r_s}{r} \right]^{1/3},
\end{eqnarray}
where $T_\bg = T(r \rightarrow \infty)$ is the background temperature.
Here,
we approximate that the freedom of the massless particles ${g_*}$ is constant.
%
The solution $T(r)$ is a decreasing function of $r$
and its slope is essentially $T(r) \propto r^{-1/3}$.
When the Hawking radiated particles have no interactions,
we expect a density distribution $\rho \propto r^{-2}$.
However, that of our solution is $\rho \propto T^4 \propto r^{-4/3}$.
The interactions in the radiated particles have an effect of
keeping their heat from escaping,
therefore the density slope of our solution
is gentler than that in the case of no interactions.

Now, we can determine the minimal thermalized radius $r_s$
and the initial temperature $T_s$ by a boundary condition:
the total energy flux of the Hawking radiation ${\cal F}_\BH$ must equal to
the total flux of out-going diffusion energy ${\cal F}$.
Our stationary solution has total out-going energy flux
\begin{eqnarray}
 {\cal F}
  &=&
  4\pi r^2 \times J_r
  \;=\;
  \frac{8\pi^3}{135}\beta_s\beta_\SM c_s {g_{*\SM}}T_s^2
\end{eqnarray}
which is, of course, $r$-independent.
We have defined $c_s = 1 - (T_\bg/T_s)^3$
as a factor of the background correction.
The total flux by the Hawking radiation
is given by Stefan-Boltzmann law:
\begin{eqnarray}
 \cal{F_\BH}
  &=&
  4\pi r_\BH^2 \times \frac{\pi^2}{120} {g_{*\SM}}T_\BH^4.
\end{eqnarray}
The boundary condition ${\cal F}_\BH = {\cal F}$
leads us to the radius and the temperature of the minimum thermalized sphere:
\begin{eqnarray}
 r_s
  &=&
  \frac{64\pi^2}{3} \sqrt{\beta_s^3 \beta_\SM c_s} \: r_\BH
  \;\simeq\; 6.6\times10^4 r_\BH,\\
 T_s
  &=&
  \frac{3}{16\pi} \frac{1}{\sqrt{\beta_s\beta_\SM c_s}} \: T_\BH
  \;\simeq\; T_\BH / 530,
\end{eqnarray}
where we have used a relation $c_s \simeq 1$
as a presupposition for later discussion.
Finally we obtain the spherical thermal distribution
surrounding the black hole:
\begin{eqnarray}
T(r) &=& \left[
	  T_\bg^3 + \frac{9}{256\pi^2}\frac{1}{\beta_\SM}\frac{T_\BH^2}{r}
	 \right]^{1/3}
\end{eqnarray}
for $r > r_s$.
We also obtain
a mean velocity of the out-going diffusing particles at radius $r$,
namely, a velocity for the out-going plasma flow:
\begin{eqnarray}
 v(r)
  &\simeq& \frac{J_r(r)}{\rho(r)}
  \;=\;
  \left(\frac{32\pi}{9}\right)^2 \beta_\SM^2
  \left[1-\left(\frac{T_\bg}{T(r)}\right)^3\right]^2
  \left(\frac{T(r)}{T_\BH}\right)^2
\end{eqnarray}
The existence of the temperature distribution $T(r)$
and the plasma flow $v(r)$
is obvious evidence that the system surrounding the black hole
is departure from the thermal equilibrium.


Let us next discuss formation of the domain wall surrounding a black hole.
The local temperature $T(r)$ is a decreasing function of $r$.
When there exist a region such that $T(r) > T_\weak$,
the local temperature configuration allows us
a nontrivial phase structure of the symmetry breaking depending on the space,
i.e., the vacuum expectation value (VEV) of Higgs doublet(s)
depends on the distance from the center of the black hole $r$.
More concretely,
the neighborhood of the black hole is symmetric phase
because of the heating-up by the black hole,
however, the region distant from the black hole is still broken phase.
Therefore,
we should find a electroweak domain wall surrounding the black hole.
Our conclusion that the wall surrounding the black hole exists
does not depend on the details of the phase transition
such as a order of the phase transition.

By the LTE approximation,
the space dependence of the Higgs doublets may take the form
\begin{eqnarray}
 \langle\phi_i(r)\rangle &=& \langle\phi_i\rangle_{T=T(r)}.
\end{eqnarray}
For simplicity,
we assume that the electroweak phase transition is the second order
and assume the simplest form of the Higgs VEV as
\begin{eqnarray}
 \left(\frac{|\langle\phi\rangle_T|}{v}\right)^2
  +
  \left(\frac{T}{T_\weak}\right)^2 &=& 1.
\end{eqnarray}
Then the Higgs VEV should be be written as
\begin{eqnarray}
 |\langle\phi(r)\rangle|
  &=&
  v f(r), \label{VEV.eq}
\end{eqnarray}
where
\begin{eqnarray}
 f(r) &=& 
  \left\{
   \begin{array}{lcl}
    0 & & (r \leq r_\DW) \\
    \sqrt{1 - \left(\frac{T(r)}{T_\weak}\right)^2} & & (r > r_\DW)
   \end{array}
  \right., \label{FormFunc.eq}
\end{eqnarray}
is a form-function of the wall and has a value from zero to one
(see \fig{BHWall.eps}).
In this configuration of the Higgs VEV,
the width of our domain wall $d_\DW$
is approximately equal to the radius of the symmetric region $r_\DW$.
By $T(r_\DW) = T_{\weak}$, we find
\begin{eqnarray}
 d_\DW \simeq r_\DW
  &=& \frac{9}{256 \pi^2}
      \frac{1}{\beta_\SM c_\weak} \frac{T_\BH^2}{T_\weak^3},
\end{eqnarray}
where $c_\weak = 1 - (T_\bg/T_\weak)^3$ is other background correction.
We see that the electroweak phase structure is determined by
the thermal structure of the black hole.
We illustrate its width with other parameters in \fig{BHTemp.eps}.


We will discuss the validity condition for the domain wall.
Our stationary LTE approximation for the wall is valid
when the size of the wall is greater than the MFP
and when lifetime of the black-hole is large enough
to keep the stationary electroweak domain wall.
The first condition is
\begin{eqnarray}
 1
  &\ll&
  d_\DW/\lambda(T_\weak)
  \;=\;
  \left(
   \frac{3}{16\pi}\frac{1}{\sqrt{\beta_s\beta_\SM c_\weak}}
   \frac{T_\BH}{T_\weak}
   \right)^2,
\end{eqnarray}
and hence we have $T_\BH \gnear 5.3\times10^4\GeV$.
The equivalent condition is $m_\BH \lnear 2.0\times10^5\kg$.
Because 
the time to construct the stationary electroweak domain wall is
\begin{eqnarray}
 \tau_\DW \simeq r_\DW / v_\DW
 &=&
 \frac{729}{262144 \pi^4}
 \frac{1}{\beta_\SM^3 c_\weak^3}
 \frac{T_\BH^4}{T_\weak^5},
\end{eqnarray}
the second condition leads us to
\begin{eqnarray}
 1
  &\ll&
  \tau_\BH / \tau_\DW
  \;\simeq\;
  \frac{5242880 \pi^2}{729} \frac{\beta_\SM^3 c_\weak^3}{g_*}
  \frac{m_\planck^2 T_\weak^5}{T_\BH^7}.
\end{eqnarray}
If we consider the black hole in the vacuum,
namely, when $T_\bg \ll T_\weak$ and $c_\weak = 1$,
we obtain the stationary-wall condition
for the Hawking temperature as
$T_\BH \lnear 1.4 \times 10^8 \GeV$.
The equivalent condition is $m_\BH \gnear 76\kg$.

In the case of the primordial black hole in the early universe
which will be discussed in the later section,
we should consider the temperature of the background universe $T_\bg$
and a non-trivial value of $c_\weak$
because
the universe is reheated up by the primordial black holes
and the universe containing the black holes is not cold.
For the primordial black holes,
we obtain the stationary-wall-conditions as 
$T_\BH \lnear 3.85 \times 10^7 \GeV$.
The equivalent condition is $m_\BH \gnear 275\kg$.
To obtain this restriction,
we have used the later relation
between the Hawking temperature and the universe temperature.
We note that we have $T_\bg \simeq 98\GeV$,
when $T_\BH = 3.85 \times 10^7 \GeV$.

When these two conditions are satisfied,
we can easily check the thermalization of the domain wall
in the meaning of the time scale.
We illustrate these conditions
by the mark with ``{\sf Thermalized Wall}'' in \fig{BHTemp.eps}.
Finally, we conclude that
the electroweak domain wall surrounding a black hole exists,
when mass of the black hole satisfies 
$76\kg \lnear m_\BH \lnear 2.0\times10^5\kg$,
as a general property of a black hole and the Standard Model.

At the end of this section,
we will compare our domain wall and
a domain wall in the ordinary electroweak baryogenesis scenarios
(the CKN model) \cite{CKN}.
In the CKN model, the structure of the domain wall 
should be determined by the dynamics of the electroweak phase transition
and the form function of the wall is assumed
\begin{eqnarray}
 f(z) &=& \frac{1}{2}[\tanh{(z/\delta)} + 1], \label{FormFuncCKN.eq}
\end{eqnarray}
where $z$ is a perpendicular direction to the wall and
$\delta$ is a width of the wall.
On the other hand,
the structure of our domain wall is determined by
the thermal structure of the black hole
and the form-function of our wall
is derived in the equation (\ref{FormFunc.eq}).
In the CKN model,
a moving domain wall with a velocity $v$
during the phase transition process of the universe
was considered and
the moving of the wall in the stationary plasma
produced a non-equilibrium effect.
In our model,
the wall keeps his position,
however,
the plasma radiated from the black hole
is flowing from the symmetric region to the broken one
with velocity at the wall:
\begin{eqnarray}
 v_\DW
  &=&
  v(r_\DW)
  \;\simeq\;
  \frac{1024 \pi^2}{81} \beta_\SM^2 c_\weak^2
  \left(\frac{T_\weak}{T_\BH}\right)^2.
\end{eqnarray}
In both cases, the plasma and domain wall moves relatively.
Therefore,
at the co-moving frame of the plasma,
we cannot distinguish the situations in the models.

\section{Baryon Number Creation by a Black Hole}\label{Baryon.sec}


We will discuss a mechanism for the baryon number creation
by the Hawking radiation
as a application of the electroweak domain wall which we have found.
We consider the Standard Model with two Higgs extension (2HSM)
for a origin of a CP-broken phase
as following the electroweak baryogenesis scenarios
proposed by Cohen, Kaplan and Nelson (CKN model) \cite{CKN}.
In the 2HSM, the electroweak domain wall can have a CP broken phase
and it is described by a un-erasable phase of the Higgs VEV.
Following the CKN model,
the CP-phased neutral Higgs VEV may be written as
\begin{eqnarray}
 \langle\phi_1^0(r)\rangle
  &=&
  v_1 f(r) \; e^{-i \Delta\theta_\CP (1-f(r))}, \label{VEVCP.eq}
\end{eqnarray}
where $f(r)$ is the form function given by equation (\ref{FormFunc.eq})
and $\Delta\theta_\CP$ is a CP-broken phase in the wall.
The difference between the wall in the CKN model and our wall
is only a difference of the form functions
in the equations (\ref{FormFunc.eq}) and (\ref{FormFuncCKN.eq}).

Here, we clarify how to satisfy
the Sakharov's three criteria for the baryon number creation \cite{Sakharov}
by our model with comparing the CKN model.
\begin{enumerate}
 \item {\it The baryon number violation}:
       In the case of both the CKN model and our model,
       the sphaleron process \cite{Sph} takes place
       in the symmetric phase region and
       the near symmetric phase region in the domain wall.
 \item {\it The C-asymmetry}:
       In the both model,
       the (extended) SM is the chiral theory.\\
       {\it The CP-asymmetry}:
       In the both model,
       an extension of the Higgs sector in the SM is assumed,
       since the domain wall has CP breaking phase.
       The simplest assumption is assuming the 2HSM.
 \item {\it Out of equilibrium}:
       In the CKN model,
       the interaction between
       the moving domain wall and the stationary plasma
       is a non-equilibrium process.
       In our model,
       the interaction between
       the stationary spherical domain wall and
       the out-going plasma flowing through the wall
       is a non-equilibrium process.
\end{enumerate}
The difference of the models is essentially how to make a out of equilibrium.
In the CKN model,
the first order phase transition of electroweak theory
is assumed since the moving domain wall 
in the universe at the critical temperature
plays a crucial role.
In our model,
the first order phase transition is not necessary
to create the domain wall,
however, the existence of the black hole is required.

The width of our domain wall $d_\DW$
is greater than the MFP of the quarks $\lambda_s(T_\weak)$,
therefore,
our domain wall belongs to a kind of ``thick wall''
which was defined by Cohen-Kaplan-Nelson \cite{CKN}.
We can consider 
a variant of spontaneous baryogenesis scenario \cite{CKN},
which we call the ``thick-wall black-hole baryogenesis''.

\begin{figure}[htbp]%
\begin{center}%
 \includegraphics{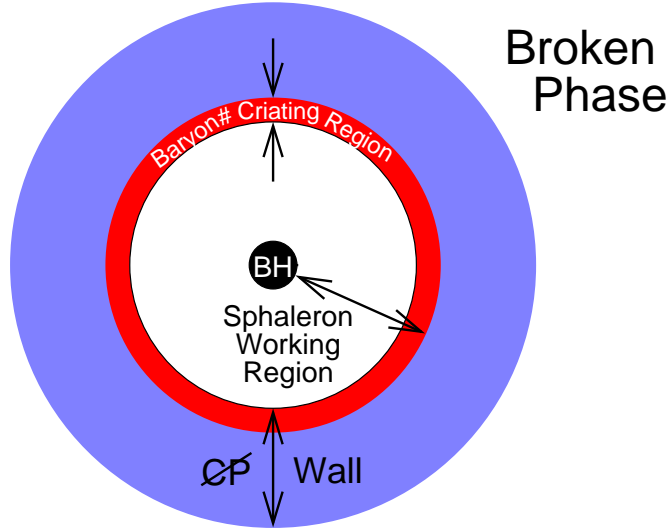}%
 \caption{%
 Baryon Number Creating region in the domain wall.
 The region is a spherical thin layer which is a intersection of
 the sphaleron working region
 and the domain wall with a CP broken phase.
 The sphaleron working region is given by
 the condition:
 $|\langle\phi(r)\rangle|/v_1 < \epsilon \simeq 1/100$.
 }
 \label{ThickWall.eps}%
\end{center}%
\end{figure}

The CP-asymmetry takes place in the domain wall
as the space-dependent physical CP phase
which is described in equation (\ref{VEVCP.eq}).
The sphaleron process, which violates the baryon number,
effectively works on the region where the Higgs VEV is small,
then the baryon-number-violating process occurs
not only in the symmetric region
but also in the domain wall near the symmetric region.
In the theory of baryon number creation with thick wall,
the baryon number is creating
at the region where all of the Sakharov's criteria are satisfied,
then the region creating the baryon number is
the spherical layer near the symmetric region in the wall
(see figure \ref{ThickWall.eps}).

Now,
we calculate how much baryon number is created
by the black hole as a function of the mass of the black hole $m_\BH$
and the CP broken phase in the wall $\Delta\theta_\CP$.
The sphaleron transition rate at the temperature $T$,
which is first obtained by Kuzmin, Rubakov and Shaposhnikov \cite{Sph}
and which is improved by Arnold, Son and Yaffe \cite{SphRate1},
is
\begin{eqnarray}
 \Gamma_\sph &=& \kappa \alpha_\weak^5 T_\weak^4 e^{-E_\sph/T_\weak},\\
\end{eqnarray}
where $\kappa \sim O(30)$ is a numerical constant \cite{SphRate2} and 
\begin{eqnarray}
  E_\sph(T) &=& \frac{2M_W(T)\:b}{\alpha_\weak}
   \;\simeq\;
  \frac{|\langle\phi\rangle|_T}{v_1} \times 10 \TeV,
\end{eqnarray}
is the energy of the sphaleron potential.
When $|\langle\phi(r)\rangle|/v_1 < \epsilon \simeq 1/100$,
namely, $E_\sph < T_\weak$,
the sphaleron transition is not suppressed by the exponential factor.
Therefore the sphaleron working region
is given by $f(r) < \epsilon$
with the equation (\ref{VEV.eq}).

To calculate the baryon number creation rate,
we need to consider the sphaleron process
only in the neighborhood of the symmetric region in the domain wall.
We define the width of this region $d_\sph$
by $f(r_\DW+d_\sph) = \epsilon$ because of the form of the Higgs VEV.
The volume integral for the baryon-number-creating-region is given by
\begin{eqnarray}
 V &=& 4\pi r_\DW^2 \times \int_{r_\DW}^{r_\DW+d_\sph} dr.
  \label{VolInt.eq}
\end{eqnarray}
In the region, we have a space-dependent CP phase $\theta(r)$
and diffusing particles have an out-going mean velocity $v_\DW$.
Then at the co-moving frame of this plasma-flow,
these particles feel the time-dependence of the CP phase:
\begin{eqnarray}
 \dot{\theta} &\simeq& v_\DW \frac{d}{dr} \theta.
\end{eqnarray}
In the ordinary spontaneous baryogenesis scenario,
the plasma containing top quarks is at rest but the domain wall is moving,
while in our scenario,
the domain wall is rest but the plasma is flowing though the domain wall.
The relation between the baryon-number chemical potential
and the time-dependent CP phase is
\begin{eqnarray}
 \mu_B &=& {\cal N} \dot{\theta},
\end{eqnarray}
where ${\cal N} \sim O(1)$ is a model-dependent constant \cite{CKN}.
Finally, we can write down
by the detailed-balance relation
the rate of the baryon number creation per black hole:
\begin{eqnarray}
 \dot{B}
 &=& - V \; \frac{\Gamma_\sph}{T_\weak} \; \mu_B
       \nonumber\\
 &=& - 4\pi {\cal N} \kappa \alpha_\weak^5 T_\weak^3 \;
       r_\DW^2 \; v_\DW
       \int_{r_\DW}^{r_\DW+d_\sph} dr \: \frac{d}{dr} \theta(r)
       \nonumber\\
 &=& - \frac{1}{16\pi} \: {\cal N} \kappa \alpha_\weak^5 \:
       \epsilon\Delta\theta_\CP \:
       \frac{T_\BH^2}{T_\weak}, \label{BRate.eq}
\end{eqnarray}
where we have used the relation
\begin{eqnarray}
 \int_{r_\DW}^{r_\DW+d_\sph} dr \frac{d}{dr} \theta(r)
  &=&
  \epsilon\Delta\theta_\CP. \label{surface.eq}
\end{eqnarray}
The total baryon number created by a black hole with a mass $m_\BH$
is given by integrating the rate in equation (\ref{BRate.eq})
over the lifetime of the black hole.
The time evolution of the Hawking temperature is
\begin{eqnarray}
 T(t) &=& \left(\frac{g_*\pi^2}{20}\right)^{1/3}
  \frac{m_\planck^{2/3}}{(\tau_\BH-t)^{1/3}},
  \label{TimeEvolution.eq}
\end{eqnarray}
where $\tau_\BH$ is a lifetime of the black hole with temperature $T_\BH$
and which is parameterized such that
$T(0) = T_\BH$ and the time $t=\tau_\BH$ is the end of the evaporation.
Finally, we obtain
the total baryon number created by a black hole in his lifetime:
\begin{eqnarray}
 B &=& \int_0^{\tau_\BH} \: dt \: \dot{B}(T(t)) \nonumber\\
   &=& - \frac{30}{\pi^2 g_*} \: {\cal N} \kappa \alpha_\weak^5 \:
       \epsilon\Delta\theta_\CP \:
       \frac{m_\BH}{T_\weak}
   \;\simeq\; 10^{-9} \times \Delta\theta_\CP \frac{m_\BH}{T_\weak}.
   \label{B.eq}
\end{eqnarray}
This result almost does not depend on the form function of the wall $f(r)$
because the volume integral in equation (\ref{VolInt.eq})
is replaced with the surface value by the equation (\ref{surface.eq}).
Indeed, the result has no parameters like $\beta$ and $T_\bg$.
The result essentially depends on the CP-broken phase $\Delta\theta_\CP$
and the mass of the black hole $m_\BH$.
We can understand that
the contribution from the phase transition dynamics
is condensed to the value of the CP-broken phase 
and the contribution from the black hole dynamics
is condensed to the mass of the black hole.

\section{Cosmology by Black Holes and Baryogenesis}\label{Cosmology.sec}

In the previous section, we have discussed that
a black hole can create a baryon number
and the producted baryon number is proportional to the mass
of the black hole.
In this section, we apply this phenomena to
the early universe, i.e.,
we will construct a baryogenesis scenario by the primordial black holes.
We will consider a scenario of the following three steps
(see Figure \ref{DHistory.eps}):
First, in the very early universe,
most of the matter existed as the primordial black holes
with its mass $m_\BH$,
i.e., the universe was black-hole-dominant.
Second,
the black holes evaporated through creating baryons in our processes.
Finally, after evaporation of these black holes,
the universe became radiation dominant at the temperature $T_\ReHeat$
by the Hawking radiation from these black holes,
i.e., the universe was reheated by the black holes.
The essence of our assumption is
the existence of black-holes-dominant in early universe,
we will discuss the possibility of the assumption later.
We have assumed the monochrome mass-spectrum of black holes
in our scenario, this is only for simplicity of calculation.
We show the procedure for the calculations given in this section
by \fig{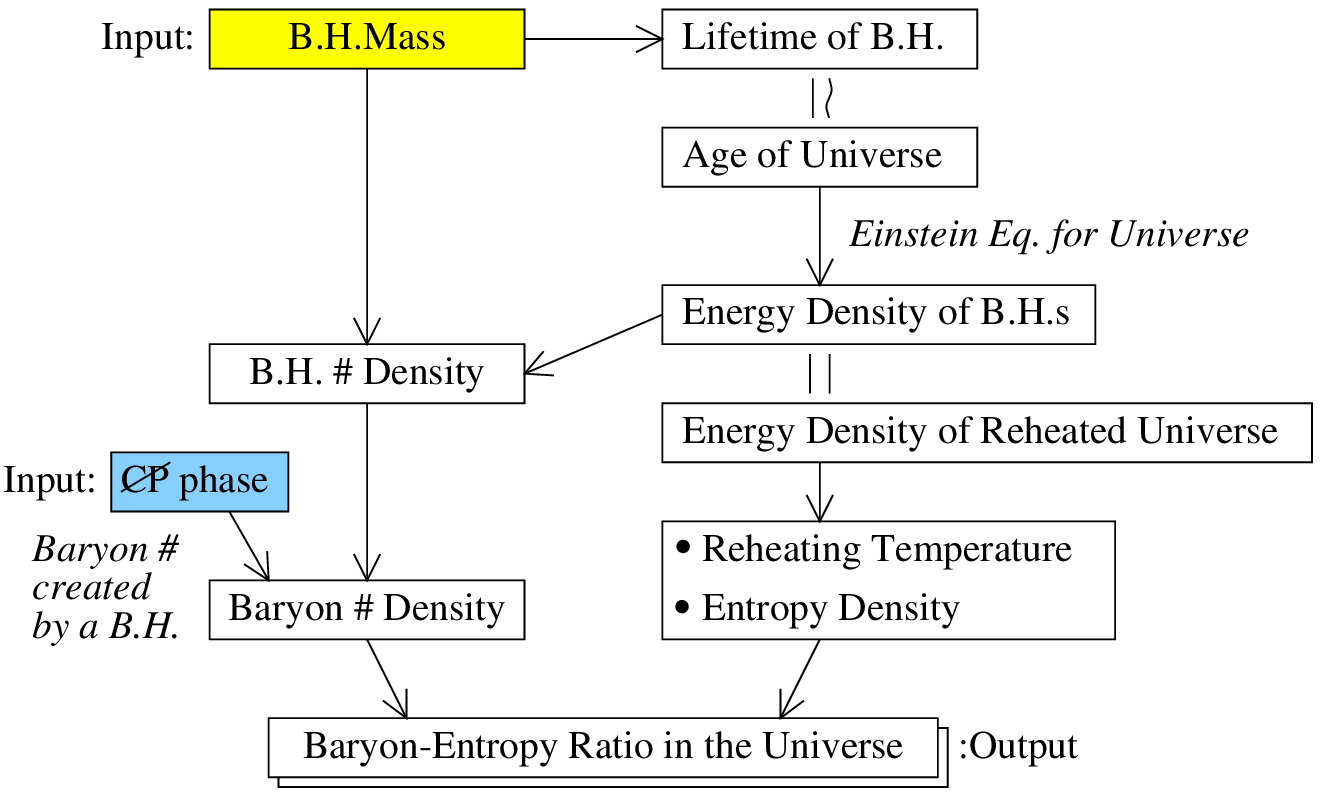} schematically.

\begin{figure}[htbp]%
\begin{center}%
 \includegraphics{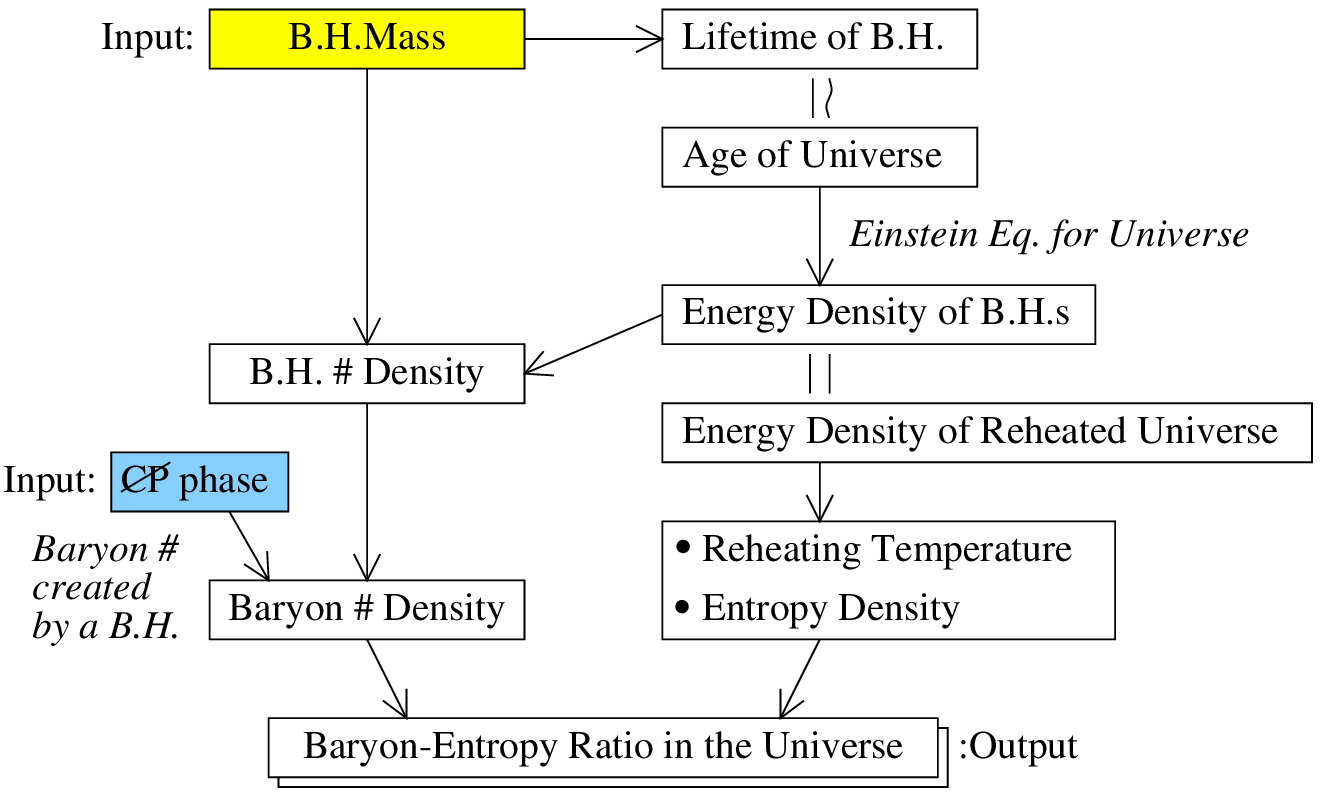}%
 \caption{
 Calculation steps from the mass of the primordial black holes and
 the CP broken phase in the wall to the resultant baryon-entropy ratio.
 }%
 \label{CalcProc.eps}%
\end{center}%
\end{figure}

The scenario implies that age of the universe $t_\ReHeat$,
when the universe was reheated by the primordial black holes 
with creating the baryon number,
equals to lifetime of the black holes:
\begin{eqnarray}
 t_\ReHeat &\simeq& \tau_\BH. \label{time-time.eq}
\end{eqnarray}
Further more, it implies
the energy density of the black holes
in the black-hole-dominant-universe $\rho_\BH$
is transfered to
one of the radiation in the radiation-dominant-universe:
\begin{eqnarray}
 \rho_\ReHeat &\simeq& \rho_\BH  \label{rho-rho.eq}
\end{eqnarray}
by the reheating process with the Hawking radiation at the time $t_\ReHeat$.

To calculate the energy density of the black-hole-dominant-universe,
we can use the Einstein equation for flat universe
with a matter dominant:
\begin{eqnarray}
 \rho_\BH &=& \rho_{\rm matter dom.}(t_\ReHeat)
  \;\equiv\; \frac{1}{6\pi} \frac{m_\planck^2}{t_\ReHeat^2} \nonumber\\
  &\simeq& \frac{\pi^3 g_*^2}{2400} \frac{T_\BH^6}{m_\planck^2}
   \label{rhoBH.eq}
\end{eqnarray}
because the universe after inflation is flat
and the Einstein equation for the black-hole-dominant-universe
is same as the one of the matter-dominant-universe
\cite{Barrow}\cite{Majumdar}\cite{KolbTurner}.
In the last equation, we have used the equation (\ref{time-time.eq})
and the expression of $\tau_\BH$.

By using the thermodynamical relation
between the density and the temperature for the radiation:
$\rho = \frac{\pi^2}{30} g_{*} T^4$
and equations (\ref{rhoBH.eq})(\ref{rho-rho.eq}),
we have a relation between
the reheating temperature and the Hawking temperature of the black holes:
\begin{eqnarray}
 g_{*\ReHeat}T_\ReHeat^4
  &\simeq& \frac{\pi g_{*\BH}^2}{80} \frac{T_\BH^6}{m_\planck^2},
  \label{eqTunivTbh}
\end{eqnarray}

In our scenario,
the evaporation of the black holes in the broken phase vacuum
is essential to create the baryon number 
and to conserve the created baryon number,
then the reheating temperature of universe must be
lower than the electroweak critical temperature $T_\weak$.
By the relation (\ref{eqTunivTbh}),
this restriction gives us
an upper bound for the black-hole temperature
and an equivalent condition for the mass of black hole as a lower bound:
\begin{eqnarray}
 T_\BH &<& T_{c\BH},
  \qquad
 m_\BH \;>\; m_{c\BH},
\end{eqnarray}
where we have defined
a critical temperature and mass for black hole:
\begin{eqnarray}
 T_{c\BH}
  &\equiv&
  \left(\frac{80}{\pi}\frac{g_{*\weak}}{g_{*c\BH}^2}\right)^{1/6}
  T_\weak^{2/3} m_\planck^{1/3}
  \;\simeq\; 3.9 \times 10^7 \GeV,\\
 m_{c\BH}
  &\equiv&
  \frac{1}{8\pi}
  \left(\frac{\pi}{80}\frac{g_{*c\BH}^2}{g_{*\weak}}\right)^{1/6}
  \frac{m_\planck^{5/3}}{T_\weak^{2/3}}
  \;\simeq\;
  \frac{1}{20.} \left(\frac{m_\planck}{T_\weak}\right)^{2/3} m_\planck
  \;\simeq\; 270\kg,
  \label{McBH.eq}
\end{eqnarray}
as the temperature of the reheated universe $T_\ReHeat(m_{c\BH})$ equals
to the critical temperature of the electroweak theory $T_\weak$.
We have another cosmological restriction;
The evaporation of the black holes should not affect
the very successful big-bang-nucleosynthesis (BBN) theory \cite{KolbTurner}.
Therefore the reheating temperature $T_\ReHeat$ must be higher
than the temperature for the BBN working
$T_\BBN \simeq 10 \MeV \sim 0.1 \MeV$,
namely,
we have a lower bound for the black-hole temperature
$T_\BH \gnear 5.9 \times 10^4 \GeV$
(see \fig{BHTemp.eps}).


When the black holes evaporate,
the number-density of black holes is given by
\begin{eqnarray}
 n_\BH &=& \frac{\rho_\BH}{m_\BH}
 \;=\; \frac{\pi^4 g_{*\BH}^2}{300}  \frac{T_\BH^7}{m_\planck^4},
\end{eqnarray}
because we have assumed the monochrome mass-spectrum of the black holes.
We already have known how much baryon number is created by a black hole
as a function of its mass or temperature and the CP broken phase, hence
the baryon number density created by our mechanism is
\begin{eqnarray}
 b &=& B \: n_\BH.
\end{eqnarray}
At the same time,
the entropy density in the universe reheated by the black holes
is given by the thermodynamical relation:
\begin{eqnarray}
 s &=& \frac{2\pi^2}{45} g_{*\ReHeat} T_\ReHeat^3.
\end{eqnarray}
By using relation (\ref{eqTunivTbh}),
we have the baryon-entropy ratios in the universe:
\begin{eqnarray}
 \frac{b}{s}
  &\simeq&
  \frac{45}{2 \pi^2 g_*} \;
  {\cal N} \kappa\alpha_\weak^5
  \epsilon \Delta\theta_\CP \:
  \frac{T_\ReHeat}{T_\weak}. \nonumber\\
 &\simeq&
  7 \times 10^{-10} \times \frac{\Delta\theta_\CP}{\pi}
  \: \frac{T_\ReHeat}{T_\weak}, \label{bsratio.eq}
\end{eqnarray}
when $3.9 \times 10^7 \GeV \gnear T_\BH \gnear 5.9 \times10^4 \GeV$.
We display this result in \fig{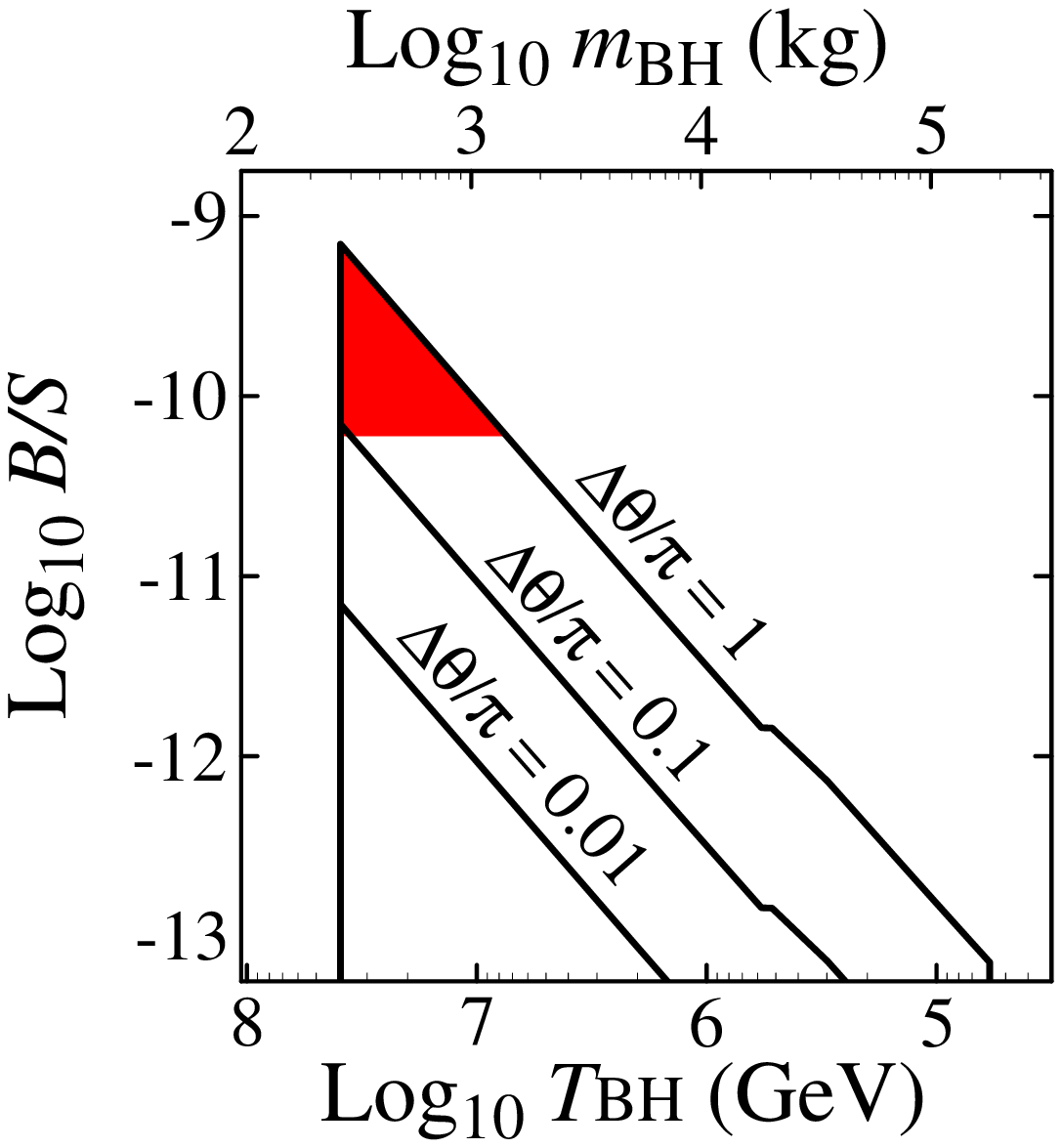}.

\begin{figure}[htbp]%
\begin{center}%
 \ \includegraphics{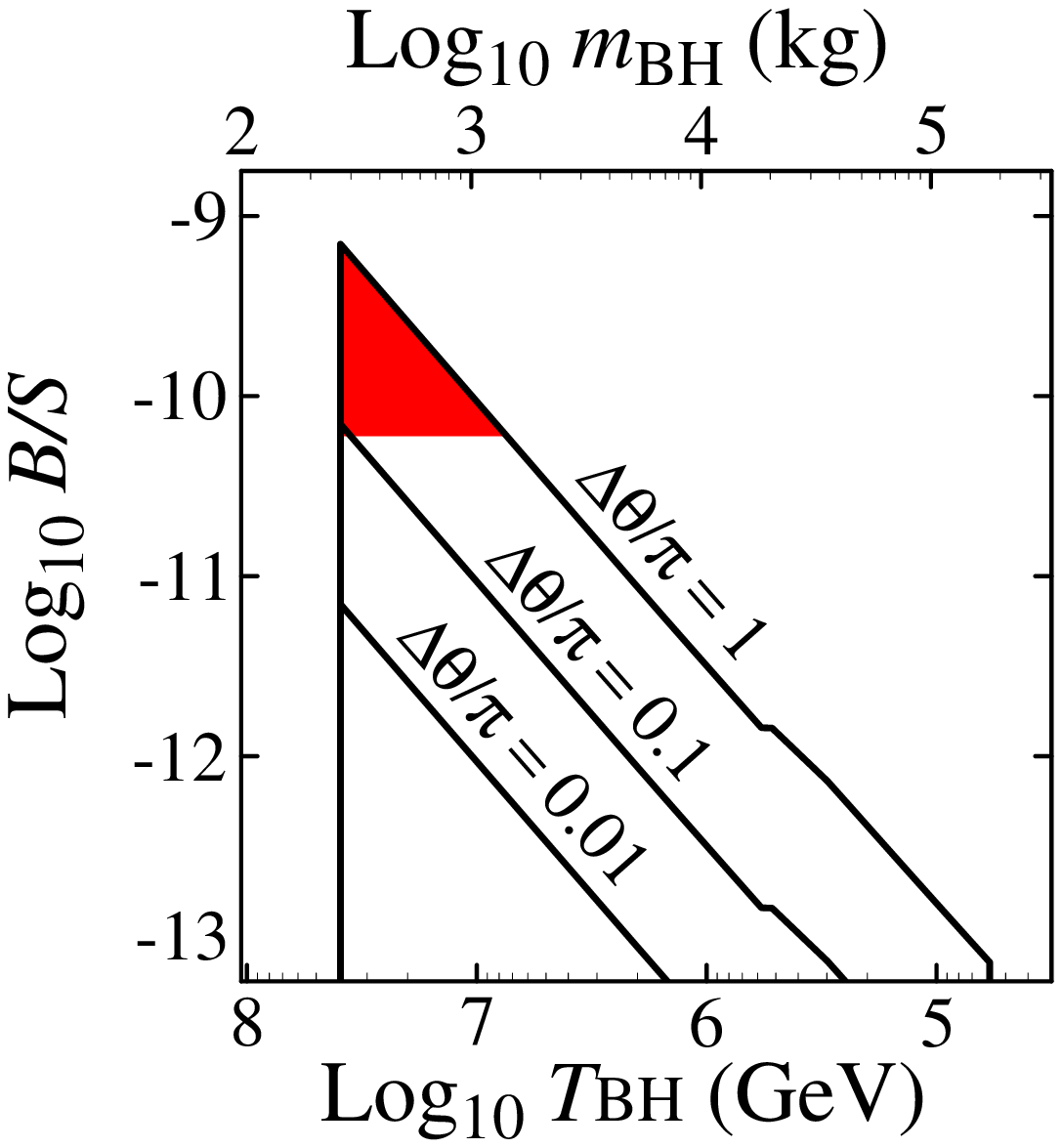}
 \caption{The resultant baryon-entropy ratio by our scenario
 	  when $\Delta\theta_\CP/\pi=1,0.1,0.01$.
 	  The shaded region satisfies the BBN requirement.
          }%
          \label{Result.eps}%
\end{center}%
\end{figure}

Finally,
if the domain wall has a CP broken phase: $\Delta\theta_\CP = \pi$,
as the most optimistic estimation,
the baryon-entropy ratio in our scenario
satisfies the BBN requirement: $B/S \sim 10^{-10}$ \cite{KolbTurner}
when $3.9 \times 10^7 \GeV \gnear T_\BH \gnear 7.4 \times 10^6 \GeV$,
namely $270\kg \lnear m_\BH \lnear 1400\kg$
(see the BBN-allowed region in \fig{Result.eps}).
The typical mass of the black hole in the scenario
is the critical mass $m_{c\BH}$
which is defined in the equation (\ref{McBH.eq}).
By introducing a new parameterization of black hole mass:
\begin{eqnarray}
 m_\BH &=& \xi m_{c\BH}, \label{xiparam}
\end{eqnarray}
the allowed region for the parameter is
$1.0 < \xi \lnear 5.2$ and the baryon-entropy ratio in equation
(\ref{bsratio.eq}) is represented by
\begin{eqnarray}
b/s &\simeq&
  7 \times 10^{-10} \times \frac{\Delta\theta_\CP}{\pi}
  \: \xi^{-3/2}, \label{bsratio2.eq}
\end{eqnarray}
where we have used that
the degree of freedom $g_{*\BH}(T_\BH)$
can be regarded as a constant in the arrowed region.
%
%
In conclusion,
we have proposed a new scenario of the baryogenesis
which does not need the first order phase transition
but does require the primordial black holes.

\section{Remnants as Cold Dark Matter}\label{DM.sec}

In this section,
we will discuss that
the cold dark matter in our universe can be regarded as
remnants of the black holes which have created our baryon number.
We will use a assumption that
any black hole leaves a remnant after evaporation
with a mass of the Planck mass scale.
Let
\begin{eqnarray}
 m_\DM &=& \zeta \: m_\planck  \label{mDM.eq}
\end{eqnarray}
be a mass of the remnant, where $\zeta$ is constant with order one.
On the other hand,
the baryonic matter created by a black hole is given by
\begin{eqnarray}
 M_B &=& B \; m_\nucl \;=\; 10^{-11} \times \Delta\theta_\CP \: m_\BH,
\end{eqnarray}
because
any baryon number in the present universe stays in a nucleon
with a mass $m_\nucl \simeq 1 \GeV$.
In the last equation,
we have used the result of baryon number created by a black hole $B$
in the equation (\ref{B.eq}).
Mass of the black hole is restricted from our baryogenesis scenario
and the scale of mass is essentially given by the critical mass in
equation (\ref{McBH.eq}).
Therefor, we can evaluate the density ratio
between the baryonic matter and the dark matter in the universe:
\begin{eqnarray}
 \frac{\rho_\DM}{\rho_B}
  &=& \frac{m_\DM}{M_B}
 \;\simeq\; 10^{11} \times \frac{\zeta}{\Delta\theta_\CP}
       \frac{m_\planck}{m_\BH}
\end{eqnarray}
which is invariant to the expansion of universe
because there are no sources for the remnants and the baryon numbers
after the reheating by the evaporation of the black holes.
By using the $\xi$-parameterization for the black hole mass,
we obtain
\begin{eqnarray}
 \frac{\rho_\DM}{\rho_B}
  &\simeq& 10^{11} \times
   \frac{20 \zeta}{\xi\Delta\theta_\CP} 
   \left(\frac{T_\weak}{m_\planck}\right)^{2/3}
   \;\simeq\; 10 \frac{\zeta}{\xi\Delta\theta_\CP}.
\end{eqnarray}
In the last equation,
we found an accidental cancelation for large numbers
between $10^{11}$,
which is determined by the mechanism of the baryon number creation,
and the value of $(T_\weak/m_\planck)^{2/3} \simeq 0.5\times10^{-11}$.
The constants $\zeta, \xi$ and $\Delta\theta_\CP$ are order one,
then we find
\begin{eqnarray}
 \Omega_\DM/\Omega_B &=& \rho_\DM/\rho_B \;\sim\; O(10).
\end{eqnarray}
More directly, we can discuss that a black hole with a mass of $300\kg$
left a remnants with a mass of $m_\planck \simeq 20\ \mu g$
and it created a baryonic matter whose amount of the mass is $3\ \mu g$.
The resultant value of $\Omega_\DM/\Omega_B$ is consistent to the observation;
The baryon density $\Omega_B \simeq 0.02 \sim 0.05$
is given by the BBN prediction or CMB observations,
and the dark matter density $\Omega_\DM \simeq 0.3$
is expected
by a combination of
the cosmological constant
from the type Ia supernovae observation \cite{SN1a}
and the flatness of the universe
from the BOOMERanG experiment \cite{BOOMERanG}.
Then we can imply that
most of the dark matter consists of the black-hole-remnants
with a mass of Planck mass scale, i.e. few tens microgram.

Next we will evaluate the number density of the remnants
in the present universe and
will estimate the number flux of the remnants on the earth.
The number density of dark matter is given by
\begin{eqnarray}
 n_\DM
  &=& \frac{\rho_\DM}{m_\DM}
  \;=\; \frac{\Omega_\DM}{\zeta}
       \frac{\rho_\crit}{m_\planck}
  \;=\; \frac{3}{8\pi} \frac{\Omega_\DM}{\zeta} H_0^2 m_\planck \nonumber\\
  &=& 0.86 \frac{h^2 \Omega_\DM}{\zeta} \left(\frac{1}{1000\km}\right)^3.
\end{eqnarray}
When we use typical values of the parameters as
$\Omega_\DM=0.3$, $h=0.7$ and $\zeta=1$,
the number density is $n_\DM \simeq (2000\km)^{-3}$,
therefore,
there exists a remnant per a cube with side few thousands kilometer long.
We can estimate the number flux of the remnants on the earth as
\begin{eqnarray}
 {\cal F}_\DM &=& n_\DM \times v_\DM,
\end{eqnarray}
where $v_\DM$ is a mean velocity of the remnants in the earth.
The order of $v_\DM$
can be considered to be given by an orbital velocity of the sun in the galaxy,
therefore, we use $v_\DM \simeq 300\km/\sec.$
Then we have an estimated value for the flux of the remnants:
${\cal F}_\DM \simeq 1 \;/{\rm km^2 / year}$.
Their kinetic energy is
$E = \frac{1}{2}m_\DM \; v_\DM^2 \;\simeq\; 10^{13}\GeV.$
A detectability for the flux of the remnants
is depending on the properties of the remnants,
such that the electric charge, the magnetic charge and QCD charges, etc.
These properties should be determined by the quantum gravity.
If they have no charges, we cannot detect them directly,
however, the existence of them is consistent with any known restrictions.

\section{Cosmological Graviton Background from Black Holes}\label{model.sec}

Any black hole radiates any kind of particles with his Hawking
temperature, because the gravitation is a universal interaction.
Then the black holes in our model should radiate graviton also.
Because the gravitational interaction is very small,
the gravitons radiated from the black holes have no chance of the
interaction to other particles and
they are preserved in the universe,
except for the red shift effect due to the expansion of universe.
We will discuss properties of the cosmological graviton background (CGB)
whose origin is the Hawking radiation of our black holes.
When temperature of a black hole is much greater than the scale of
the Standard Model,
the total energy of each radiated particle
proportions to his degree of freedom $g_*$.
Then, just after the evaporation of black holes,
the energy density ratio between the graviton and particles in the Standard
Model is given by
\begin{eqnarray}
 \frac{\rho_\grav}{\rho_\SM} &=& \frac{g_{*\grav}}{g_{*\SM}},
\end{eqnarray}
where $g_{*\grav} = 2$ is degree of freedom of graviton and
$g_{*\SM} = 106.75$ is that of particles in the Standard Model.
The radiation energy of the Standard Model particles
is transfered to that of the photons $\rho_\gamma$ and
that of the neutrinos $\rho_\nu$.
By considering the effect of freeze out and decoupling of particles,
or by using a preservation of the entropy in a co-moving volume,
we can obtain the ratio of energy density of graviton (CGB)
to that of the photon with $2.75\K$ (CMB) in the present universe:
\begin{eqnarray}
 \frac{\rho_{\grav}}{\rho_{\gamma}} &=& \frac{g_{*\grav}}{g_{*\SM}}
  \left(\frac{g_{*\nu e\gamma}}{g_{*\SM}}\right)^{1/3}
  \frac{g_{*\nu e\gamma}}{g_{* e\gamma}}
  \left(\frac{g_{*\gamma}}{g_{*e\gamma}}\right)^{1/3}
  \;=\; \left(\frac{g_{*s}}{g_{*\SM}}\right)^{4/3}
  \;=\; \frac{1}{82.2},
\end{eqnarray}
where $g_{*\nu e\gamma} \equiv g_{*\nu} + g_{*e} +g_{*\gamma}$,
$g_{*e\gamma} \equiv g_{*e} +g_{*\gamma}$,
$g_{*\nu} = 7/8 \cdot 3 \cdot 2$,
$g_{*e} = 7/8 \cdot 2 \cdot 2$ and
$g_{*\gamma} = 2$ are degrees of the freedom of the each particles.
This result is equal to the energy-density of the CGB with $0.91$ K
by the model of {\it graviton decoupling}
from the thermal equilibrium at the Planck time \cite{KolbTurner}.
However, the spectrum of our CGB is much different from the thermal one.
A small black hole radiates the high energy free gravitons,
however,
the high energy SM-particles radiated from the black hole
are thermalized and their temperature should be lower than
the electroweak critical temperature $T_\weak$.
Therefore the energy of the each graviton is much greater than one of CMB.

An energy spectrum of the graviton flux from a black hole
with temperature $T(t)$ at a time $t$ is given by
\begin{eqnarray}
 dJ(t) &=& \frac{1}{4} d\rho(t) \times S_\BH(t)
  \;=\; \frac{g_{*\grav}}{32 \pi^3} \frac{k^3}{T(t)^2}
  \frac{dk}{e^{k/T(t)} - 1}, \label{GFluxSpec.eq}
\end{eqnarray}
where $k$ is a absolute value of the graviton momentum,
namely a energy of graviton,
and $S_\BH(t)$ is a surface area of the black hole horizon.
The spectrum of graviton in the universe
just after the evaporation of the black hole
is given by a superposition of the spectrum of the graviton flux in
equation (\ref{GFluxSpec.eq}) with varying temperature:
\begin{eqnarray}
 dE &=& \int_{0}^{\tau_\BH} dt dJ(t) \nonumber\\
 &=& \frac{g_{*\grav}}{32\pi^3} k^3 dk
  \int_{0}^{\tau_\BH} dt \frac{1}{T^2(t)} \frac{1}{e^{k/T(t)}-1}
\end{eqnarray}
By using the time evolution of the Hawking temperature
given in equation (\ref{TimeEvolution.eq}),
we have
\begin{eqnarray}
 dE &=& \frac{15}{\pi^4} \frac{g_{*\grav}}{g_*} m_\BH
   \frac{k^3 dk}{T_\BH^4} {\cal K} \left(\frac{k}{T_\BH}\right),
   \label{InitSpec.eq}
\end{eqnarray}
where we have defined a function
\begin{eqnarray}
 {\cal K}(x) &\equiv& \int_1^\infty \frac{ds}{s^6} \frac{1}{e^{x/s} - 1},
\end{eqnarray}
which has a property for the normalization:
\begin{eqnarray}
 \frac{15}{\pi^4} \int_0^\infty dx \; x^3 {\cal K}(x) &=& 1 
\end{eqnarray}
The form of the spectrum is shown in \fig{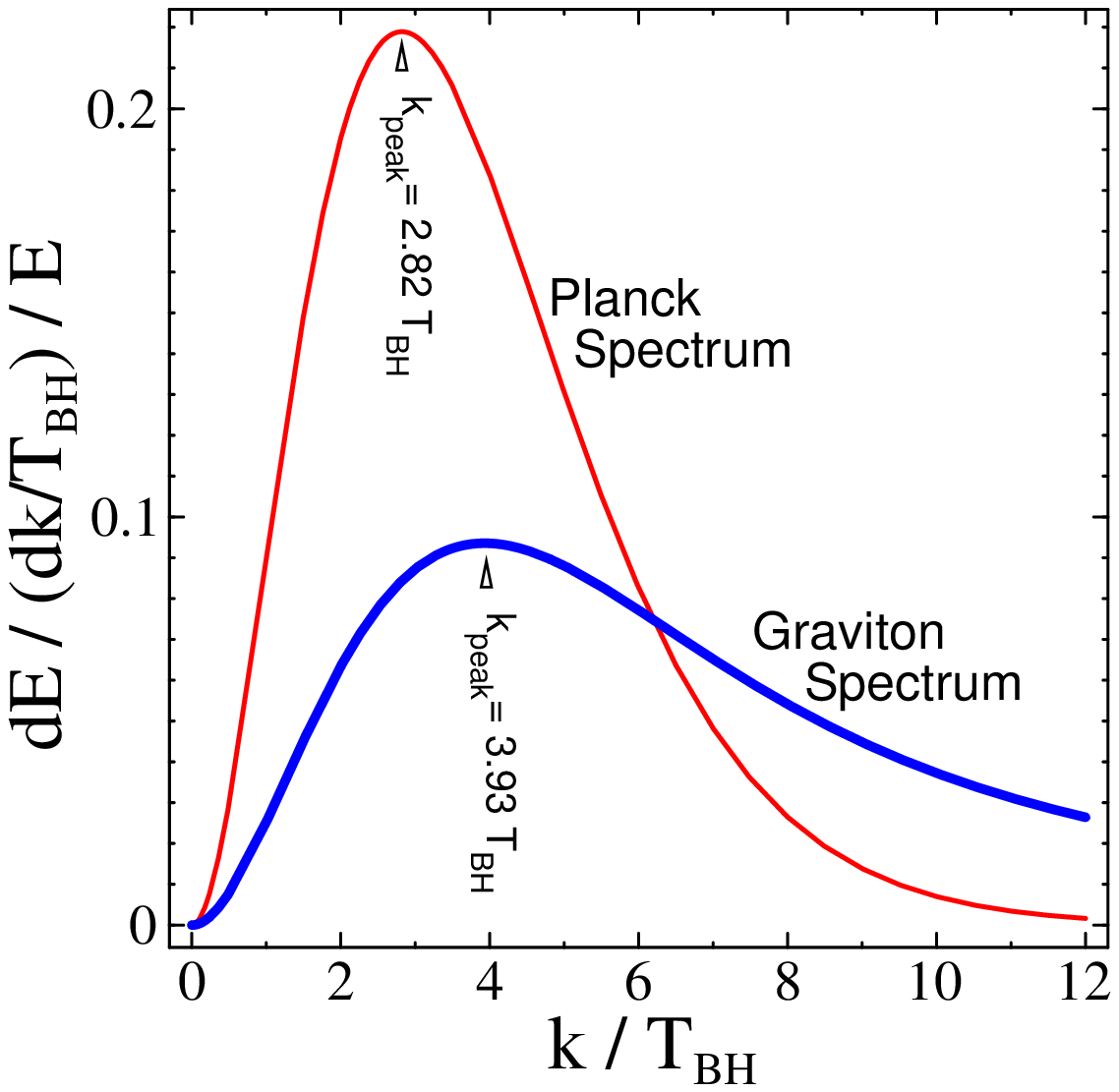}.
The resultant spectrum is different from the Planck's spectrum;
The peak of the Planck's spectrum is given by
$k_\peak = 2.82 \times T_\BH$
as known as the Wien's displacement law.
The peak of the graviton spectrum is
a little bigger than that of the Planck spectrum
because the Hawking temperature becomes higher according to the time.
We can obtain numerically 
the peak energy of the graviton spectrum
just after the evaporation of the black hole
with the initial Hawking temperature $T_\BH$:
\begin{eqnarray}
 k_{\peak,\evap} &=& 3.93 \times T_\BH. \label{Wien.eq}
\end{eqnarray}

\begin{figure}[htbp]%
\begin{center}%
 \includegraphics{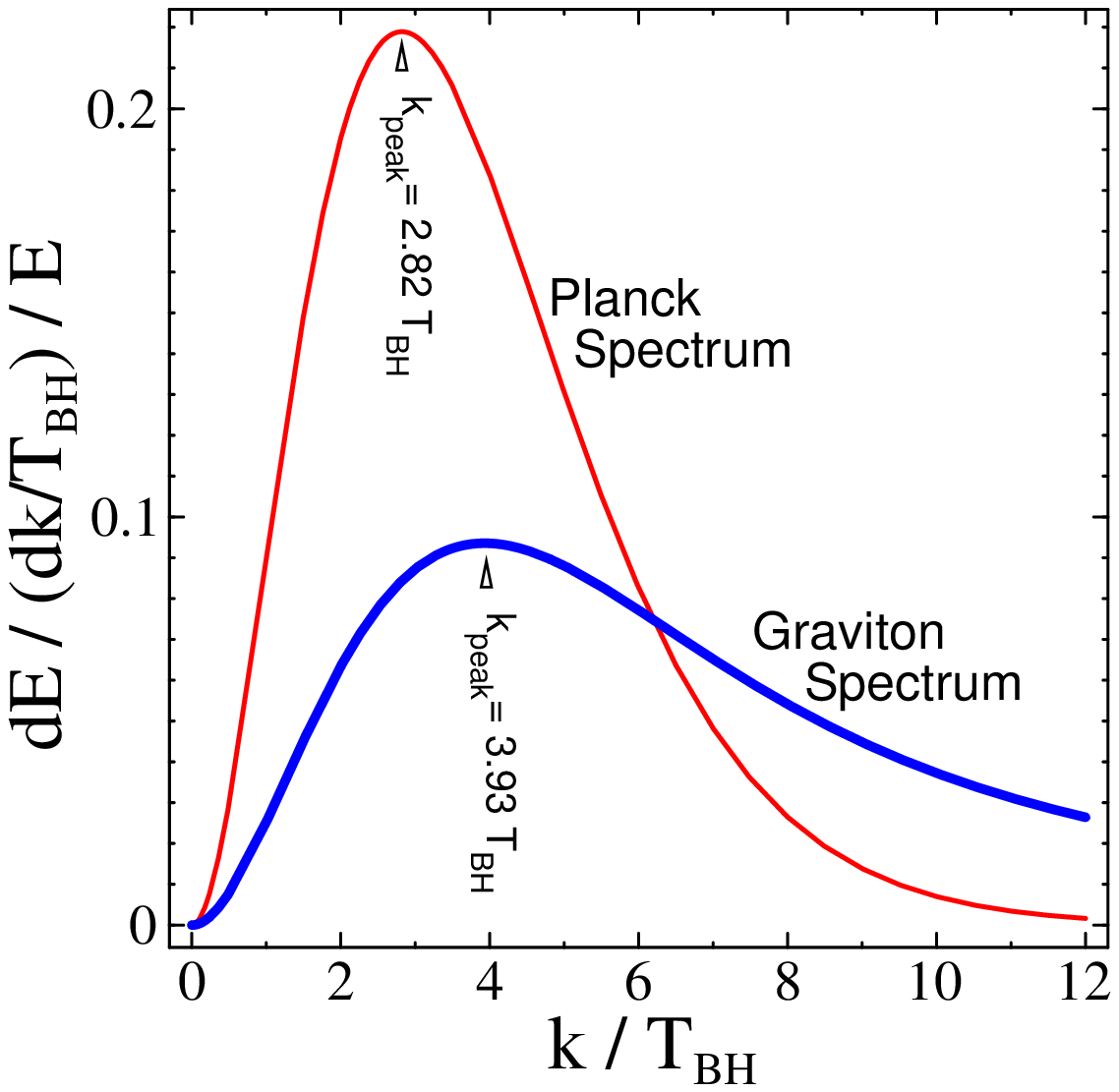}%
 \caption{
 Energy spectrum of the gravitons
 just after the evaporation of the black holes
 with initial Hawking temperature $T_\BH$.
 $E \equiv \int_0^\infty \frac{dE}{dk}dk = m_\BH \frac{g_{*\grav}}{g_{*}}$
 is a toral energy of radiated gravitons.
 Our resultant graviton spectrum is described by the thick (blue) curve.
 To compare the ordinary Planck spectrum with temperature $T_\BH$,
 we put it as the thin (red) curve.
 The peak of the Planck spectrum is given by Wien's formula:
 $k_\peak = 2.82 \times T_\BH$, however,
 the peak of our spectrum is numerically evaluated as
 $k_\peak = 3.93 \times T_\BH$.
 In the present universe,
 the peak-energy of the gravitons arises at $120\sim280\eV$
 by considering the red-shift effect from the expansion of the universe
 and their energy-density is $(g_{*s}/g_{*\SM})^{4/3}\simeq82.2$
 of the CMB-energy-density.
 }%
 \label{GravSpec.eps}%
\end{center}%
\end{figure}

To obtain the present spectrum of the CGB, we should consider
the red-shift effect by the expansion of the universe.
The red-shift factor
is given by the ratio of the scale factor of the universe,
and ratio of the scale factor is given by the co-moving entropy
preservation: $s_\ReHeat R_\ReHeat^3 =  s_{\nu\gamma} R_0^3$,
where
$s_\ReHeat =\frac{2\pi^2}{45}g_{*\SM}T_\ReHeat^3$
and
$s_0 =\frac{2\pi^2}{45}g_{*s}T_\gamma^3$.
The red-shift factor for the reheating time is
\begin{eqnarray}
 Z_\ReHeat
  &=&
  \frac{R_0}{R_\ReHeat}
  \;=\;
  \left(\frac{g_{*\ReHeat}}{g_{*s}}\right)^{1/3}
  \frac{T_\ReHeat}{T_\gamma}
  \;\simeq\; 
  \left(\frac{g_{*\weak}}{g_{*s}}\right)^{1/3}
  \frac{T_\weak}{T_\gamma} \xi^{-3/2}.
  \label{redshift.eq}
\end{eqnarray}
We can evaluate the peak energy of the CGB in the present universe
by the equations (\ref{Wien.eq}) and (\ref{redshift.eq}):
\begin{eqnarray}
 k_{\peak,0} &=&
  \frac{k_{\peak,\evap}}{Z_\ReHeat} 
  \;=\;
  3.93 \times 
  \left(\frac{g_{*s}}{g_{*\weak}}\right)^{1/3}
  \frac{T_{c\BH}}{T_\weak}
  \xi^{1/2}
  T_\gamma\\
  &\;=\;&
  3.93 \times 
  \left(\frac{g_{*s}}{g_{*\weak}}\right)^{1/3}
  \frac{20.}{8\pi}
  \left(\frac{m_\planck}{T_\weak}\right)^{1/3}
  \xi^{1/2}
  T_\gamma.
  \;=\;
  122 \; \xi^{1/2} \eV.
\end{eqnarray}
Finally, our model implies the existence of
a cosmological graviton background (CGB),
which has a characteristic energy spectrum
and its peak should be $120 \sim 280 \eV$,
and whose energy density is $1/82.2$ of CMB's.

\section{Restrictions from Formation of the Black Holes}\label{form.sec}

In this section,
we discuss conditions for the inflation model
to realize the black hole dominant universe
with a mass of several hundred kilograms.
The formation of primordial black holes by a fluctuation
from the inflation era is discussed
by Carr \cite{CarrForm, FormationByInflation}.
During the inflation era,
a density fluctuation is creating as a quantum fluctuation
and whose strength is a function of the time when the fluctuation is
created.
For example,
the fluctuation in the simplest model for the inflation is given by 
\begin{eqnarray}
 \delta(t)
 &\simeq&
 \frac{1}{m_\planck^3} \;
    \left.\frac{V^{3/2}}{V'}
 \right|_{\phi = \phi(t)},
\end{eqnarray}
where $V(\phi)$ is a inflaton potential.
The created fluctuation has once gone out to the outside of the horizon
at the inflation era,
namely, the fluctuation has frozen.
After the end of the inflation, the fluctuation reenter to the horizon.
Therefore, we can consider that the fluctuation is a function of the time
when the fluctuation reenters.
Here, we regard the strength of the fluctuation $\delta$
as a function of the reenter time $t_\reenter$.
When the fluctuation reenter to the horizon,
the fluctuation can grow into the black hole.
The formation probability for the black holes is
\begin{eqnarray}
 \beta(\delta(t_\reenter))
  &\simeq&
  \delta(t_\reenter) \;
  \exp\left( -\frac{1}{18 \delta^2(t_\reenter)} \right)
  \label{Carr.eq}
\end{eqnarray}
and the mass of the black holes is given by the horizon mass of the universe:
\begin{eqnarray}
  M_{\rm HOR}(t_\reenter)
   &\equiv&
  \frac{4\pi}{3} \;t_\reenter^3 \times \rho(t_\reenter)
  \;=\; \frac{1}{8} m_\planck^2 \; t_\reenter.
\end{eqnarray}

Our model requires black holes with a mass
\begin{eqnarray}
 m_\BH
  &=&
  \frac{\xi}{20.}\left(\frac{m_\planck}{T_\weak}\right)^{2/3} m_\planck,
\end{eqnarray}
where $1.0 < \xi \lnear 5.2$,
then the black holes should be created at the time $t_\formation$
such that $m_\BH = M_{\rm HOR}(t_\formation)$.
Here we conclude
the age of universe when black holes are created is
\begin{eqnarray}
 t_\formation &=&
  \frac{2}{5} \xi
  \left(\frac{m_\planck}{T_\weak}\right)^{2/3}
  \frac{1}{m_\planck}
  \;\simeq\; 5.3 \; \xi \times 10^{-33} \sec,
\end{eqnarray}
and the fluctuation $\delta(t)$ should have a sharp peak at $t=t_\formation$.
The temperature of the universe when the black holes are formated is
calculated from the description of $\rho(\tau_\formation)$:
\begin{eqnarray}
 T_\formation
  &=&
  0.27 \frac{1}{\sqrt{\xi}}
  \left(\frac{T_\weak}{T_\planck}\right)^{1/3}
  m_\planck
  \;\simeq\;
  6.6 \frac{1}{\sqrt{\xi}} \times 10^{12}\GeV,
\end{eqnarray}
consequently
the reheating at the end of inflation
should heat up the universe more highly than $T_\formation$.

Our model also requires black hole dominate
when the black holes evaporate and reheat the universe.
By the relation in equation (\ref{time-time.eq}),
the age of the reheating universe is
\begin{eqnarray}
 t_\ReHeat
  &=&
  \frac{\xi^3}{28.}
  \left(\frac{m_\planck}{T_\weak}\right)^2
  \frac{1}{m_\planck}
  \;\simeq\;
  2.9\;\xi^3 \times 10^{-11} \sec.
\end{eqnarray}
We require the black hole domination at $t_\ReHeat$,
not at $t_\formation$.
In the expanding universe with a scale factor $R$,
the behavior of the density of the black holes:
$\rho_\BH \propto R^{-3}$
is different from that of the radiation created by the end of the inflation:
$\rho_{\rm rad} \propto R^{-4}$ (see Figure \ref{DHistory.eps}).
Even if few black holes is created at $t_\formation$,
the black hole dominated universe can be realized before $t_\ReHeat$.
The requirement for $\beta(t_\formation)$ is
\begin{eqnarray}
 \beta(t_\formation)
  &>& \frac{R(t_\formation)}{R(t_\ReHeat)}
  \;>\; \left(\frac{t_\formation}{t_\ReHeat}\right)^{1/2}
  \;\simeq\; 10^{-11},
\end{eqnarray}
and the same condition for $\delta$ from the equation (\ref{Carr.eq}) is
\begin{eqnarray}
 \delta(t_\formation) &\gnear& 0.05.
\end{eqnarray}

\section{CONCLUSION AND DISCUSSIONS}\label{summary.sec}

%
In this paper,
we have shown a formation of the spherical electroweak domain wall
around a small black hole,
as a general property of the Hawking radiation in the standard model (SM),
and a creation of a baryon number in the wall
as an application of the wall.
The total baryon number created by a black hole
is proportional both to the mass of the black hole and
to the CP broken phase in the wall.
The property of the black hole is applicable to the cosmology.
We have proposed a model which can explain both
the origin of the baryon number and
the origin of the cold dark matter
in the universe
by the primordial black holes with a mass of several hundred kilograms
and their remnants with a Planck mass.
The model also predicts the existence of
the cosmological graviton background
whose energy spectrum is non-thermal
and has a peak at the energy $120 \sim 280 \eV$.



The formation of the electroweak domain wall
is a general property of the black hole, however,
the other results are depending on some assumptions.
Our cosmological scenario for the baryogenesis and the dark matter
requires the following three conditions;
(i) the early universe has been dominated by the primordial black holes 
with a mass of several hundred kilograms,
(ii) the electroweak domain wall has
a CP broken phase with the order of one and
(iii) any black hole leaves a remnant with a Planck mass.


In order to create the primordial black holes
with a mass of several hundred kilograms,
we have considered a density-fluctuation
whose spectrum has a sharply peak.
The simplest models of the inflation with one scalar field (inflaton)
predict almost scale invariant density-fluctuations,
then the simplest inflation models are not suitable for our purpose.
Yokoyama has discussed that
the density-fluctuation with a sharply peaked spectrum
can be created by the inflation model with multiple inflatons
and the primordial black holes with a sharply peaked mass-spectrum
are formated by the fluctuation \cite{YokoyamaPBH}.
Kawasaki {\it et al.} have discussed
the formation of the primordial black holes
with a specific mass
by the double inflation model \cite{KawasakiPBH}.
Their purpose is the explanation for 
the massive compact halo objects (MACHOs) by the primordial black holes
with a solar mass scale.
Their models may be applicable to our purpose
by choosing the parameters in their models.

Carr {\it et al.} have discussed 
the constraints for
the mass and the density of the primordial black holes
by all the results from the observations and the possible theories
\cite{CarrLimit}.
Since our primordial black holes create the baryon number,
the ``entropy constraint'' does not arise.
This fact allows us to
find a parameter window for the black-hole-dominant-universe
between the ``relics (remnants) constraint'' and the ``BBN constraint''.
The parameter window seems to be the uniquely allowed and possible region.
The primordial black holes in our model possess the parameter window.


For we needs an additional CP broken phase
to create a baryon number in the wall,
we have assumed the SM with two Higgs doublets extension (2HSM)
as a simplest extension of the SM
according to the CKN model \cite{CKN}.
After the CKN model was proposed,
many variants of the CKN model
with the minimal supersymmetric standard model (MSSM)
\cite{CKNSUSY, AokiOshimoSugamoto, Carena},
with heavy neutrinos \cite{HeavyNuBG},
with vector-like quarks \cite{VectorLikeBG} and so on,
had been proposed.
These models employ the electroweak domain wall
with the first order phase transition.
The origins of the CP broken phase in these models
may be applied to create the baryon number in our wall
and the requirement of the first order phase transition
in these models becomes unnecessary
because our domain wall is also created
even in the case of the second order phase transition.
We may also apply the diffusion enhancement effect \cite{CKN2, CKNSUSY}
to our model.


If any black hole leaves a remnant with a mass of the Planck scale,
the energy density of the remnants in the present universe
is suitable for that of the cold dark matter.
When the remnant has no charge of any kinds except for the mass,
the existence of the remnants is not restricted by the observations,
however, we may not be able to detect them directly.
In general,
there are possibilities for 
the remnant with electric charges, magnetic charges,
dipole moments, QCD charges (colors) and weak isospin.
The remnants with large positive electric charge have been excluded
by the experiments,
however,
the remnants with a few electric charge and negative electric charge
are allowed.
Considering the detectability for the remnants with charges of any kinds
would be future subject.

\begin{flushleft}
 {\Large\bf ACKNOWLEDGMENTS}
\end{flushleft}

We would like to thank
M.~Nojiri, M.~Ninomiya, K.~Kohri, T.~Sakai and K.~Shigetomi
for their useful discussions and suggestions.
We also appreciate helpful comments and advice of
A.~I.~Sanda, K.~Yamawaki, H.~Kawai, A.~Hosoya, A.~Sugamoto,
K.~Ishikawa and N.~Kawamoto.
YN is indebted to the Japan Society for
the Promotion of Science (JSPS) for its financial support.
The work is supported in part by a Grant-in-Aid for Scientific Research
from the Ministry of Education, Culture, Sports, Science and Technology
(No. 03665).

\newcommand{\PRL}[3]	{{Phys.\ Rev.\ Lett.}   {\bf #1}, #2 (#3)}
\newcommand{\PR}[3]	{{Phys.\ Rev.}          {\bf #1}, #2 (#3)}
\newcommand{\PRA}[3]	{{Phys.\ Rev.\ A}       {\bf #1}, #2 (#3)}
\newcommand{\PRD}[3]	{{Phys.\ Rev.\ D}       {\bf #1}, #2 (#3)}
\newcommand{\PL}[3]	{{Phys.\ Lett.}         {\bf #1}, #2 (#3)}
\newcommand{\PLA}[3]	{{Phys.\ Lett.\ A}      {\bf #1}, #2 (#3)}
\newcommand{\PLB}[3]	{{Phys.\ Lett.\ B}      {\bf #1}, #2 (#3)}
\newcommand{\NuP}[3]	{{Nucl.\ Phys.}         {\bf #1}, #2 (#3)}
\newcommand{\PTP}[3]	{{Prog.\ Theor.\ Phys.} {\bf #1}, #2 (#3)}
\newcommand{\Nature}[3]	{{Nature\ (London)}     {\bf #1}, #2 (#3)}
\newcommand{\PKNAW}[3]	{{Proc.\ K.\ Ned.\ Akad.\ Wet.}     {\bf #1}, #2 (#3)}
\newcommand{\Physica}[3]{{Physica\ (Utrecht)}   {\bf #1}, #2 (#3)}
\newcommand{\JMP}[3]	{{J.\ Math.\ Phys.}     {\bf #1}, #2 (#3)}
\newcommand{\PRSLA}[3]	{{Proc.\ R.\ Soc.\ London,\ Ser A}   {\bf #1}, #2 (#3)}
\newcommand{\AP}[3]	{{Ann.\ Phys.\ (N.Y.)}  {\bf #1}, #2 (#3)}
\newcommand{\JPA}[3]	{{J.\ Phys.\ A}         {\bf #1}, #2 (#3)}
\newcommand{\ZhETF}[3]	{{Zh.\ \'{E}ksp.\ Teor.\ Fiz.\ Pis'ma.\ Red.}  {\bf #1}, #2 (#3)}
\newcommand{\JETP}[3]	{{JETP\ Lett.}          {\bf #1}, #2 (#3)}
\newcommand{\CMP}[3]	{{Commun.\ Math.\ Phys.}{\bf #1}, #2 (#3)}
\newcommand{\MPLA}[3]   {{Mod.\ Phys.\ Lett.}   {\bf A#1}, #2 (#3)} 


\end{document}